\documentclass[pre,preprint,aps,superscriptaddress,showpacs,floatfix,landscape]{revtex4}
\usepackage{lscape}
\usepackage{rotating}
\usepackage{graphicx}
\usepackage{dcolumn}
\usepackage{bm}
\usepackage[usenames,dvipsnames]{color}
\usepackage{amsmath}
\usepackage{amssymb}
\usepackage{txfonts}
\usepackage{hyperref}
\begin{document}
%
%

\title{Mode instabilities and dynamic patterns in a colony of self-propelled surfactant particles covering a thin liquid layer.}

%
\author{Andrey Pototsky}
\affiliation{Department of Mathematics, Faculty of Science Engineering and Technology, Swinburne University of Technology, Hawthorn, Victoria, 3122, Australia}
\author{Uwe Thiele}
\email{u.thiele@uni-muenster.de}
\homepage{http://www.uwethiele.de}
\affiliation{Institut f\"ur Theoretische Physik, Westf\"alische
 Wilhelms-Universit\"at M\"unster, Wilhelm Klemm Str.\ 9, D-48149 M\"unster, Germany}
\affiliation{Center of Nonlinear Science (CeNoS), Westf{\"a}lische Wilhelms Universit\"at M\"unster, Corrensstr.\ 2, 48149 M\"unster, Germany}
\author{Holger Stark}
\email{Holger.Stark@tu-berlin.de}
\homepage{http://www.itp.tu-berlin.de/stark}
\affiliation{Institut f{\"u}r Theoretische Physik, Technische Universit{\"a}t Berlin, Hardenbergstrasse 36, 10623, Berlin, Germany}

\begin{abstract}
 We consider a colony of point-like 
 self-propelled surfactant particles (swimmers) 
 without direct interactions
that cover a thin liquid layer 
on a solid support. Although the particles predominantly swim 
normal
to the free film surface, their motion also has a component parallel to the film surface. The coupled dynamics of the swimmer density 
and film height profile is captured in a long-wave model allowing for diffusive and convective transport of the swimmers (including 
rotational diffusion). The dynamics of the film height profile is determined by
three physical effects:
the upward pushing force of the swimmers onto the 
liquid-gas interface that always 
destabilizes the flat film,
the solutal Marangoni force due to gradients in the swimmer concentration that always acts stabilising, and finally 
the rotational diffusion of the swimmers
together with the in-plance active motion
that 
acts either
stabilising or destabilising. After reviewing and extending the analysis of the linear stability of the flat film with 
uniform
swimmer density, we analyse the 
full
nonlinear 
dynamic equations
and show that 
point-like swimmers,
which only interact via long-wave deformations of the liquid film, 
self-organise in highly regular (standing, travelling and modulated waves) and various irregular 
patterns 
for swimmer density and film height.
\end{abstract}

\maketitle
%
%
\section{Introduction}
%
The self-assembly and self-organization of large numbers of microorganisms and artificial microswimmers and the non-equilibrium phase transitions that result from their collective behaviour have recently become the focus of many theoretical and experimental studies \cite{Rama2010arcmp,MJRL2013rmp}. Thus, in a series of experiments, carried out with different types of artificial microswimmers, several collective phenomena have been reported such as dynamic clustering, phase separation, and swarming \cite{Buttinoni_13,Theurkauff_12,Palacci_10,Palacci_13,Shashi_11}. 
In experiments with suspensions of motile living cells (e.g. {\it E.coli} and {\it B.subtilis} bacteria or spermatozoa), a variety of regular and irregular large- and meso-scale density patterns has been found \cite{Riedel05,Goldstein04,Goldstein14,sokolov07,sokolov09,liu12,sokolov09_pre,wensink_pnas,ishikawa11,dombrowski04,Schwarz-Linek13032012,Dunkel13_prl}. With the typical body size of several $\mu$m, the colonies of motile cells exhibit arrays of circular vortices, 
swirls, and meso-scale turbulence with the correlation length of the collective motion ranging between $\sim 10\mu$m and $\sim 100\mu$m. 
The emergence of large-scale coherent structures in systems composed of small-scale self-propelled objects is universal and independent of the mechanism of motility. Thus, density waves with $50-100\mu$m wavelengths are observed in an assay of $1-10\mu$m long actin filaments, driven by motor proteins \cite{Schaller10}. Stable networks of interconnected poles and asters are found in systems of microtubuli driven by kinesin complexes \cite{Surrey01}. Similar to the suspensions of sea urchins spermatozoa \cite{Riedel05}, highly coherent arrays of circular vortices were found in motile assays of microtubulus, propelled by surface-bound motor proteins \cite{Sumino_nature}. 

In order to explain the observed large- and meso-scale patterns, many theoretical models of interacting self-propelled particles have been suggested and tested against the experimental findings. 
One of the central questions of modelling is to determine the driving force and the minimal conditions for the emergence of each of the observed patterns. Historically, the first class of the developed models are the so-called {\it dry} systems, in which the motion of the embedding fluid medium is neglected \cite{Vicsek95,Bricard13,Aranson15,Grossmann14}. In contrast, in {\it wet} systems, the motion and influence of the medium is considered as well (see e.g. \cite{lauga09} and references therein).
In dry systems the formation of density patterns is triggered by a linear instability of the homogeneous isotropic state (i.e., the trivial state). The instability is caused by the direct interaction between the particles,
which
is not mediated or induced by a solvent. In the case of electrically neutral 
and
non-magnetic particles, the interaction mechanisms can be roughly divided into two categories. The first category 
deals with
steric effects such as the hard-core repulsion between colliding particles \cite{Buttinoni_13,Bialke13,Peruani06,Marchetti08,Loewen08}. 
All other 
interaction types
are due to long-range forces and fall within the second category. 
They
have to be introduced phenomenologically, such as the aligning or anti-aligning interaction in Vicsek-type flocking models, \cite{Vicsek95,Aranson15,Grossmann14}
or are due to hydrodynamic \cite{Zoettl14,Hennes14} or phoretic interactions \cite{Pohl14,Pohl15}.


The physical mechanism of 
aligning 
interactions
can be explained by 
collisions between swimmers with elongated bodies \cite{Aranson07,Marchetti08} or by the bundling of flagellas of two colliding bacteria \cite{wensink_pnas}. The true origin of 
long-range anti-aligning 
interactions
has not 
been properly explained
yet. In the case of 
{\it wet} systems, the motion of the solvent medium gives rise to hydrodynamic interactions between the suspended particles \cite{lauga09}. Several experimental and theoretical studies show that the inclusion of 
hydrodynamic interactions may destabilize the polar order at high densities, thus, it effectively acts as a long-range 
anti-aligning
force \cite{Goldstein14,Koch09}.

Presently, it is understood that the instability of a homogeneous suspensions of self-propelled particles can be induced by combining particle motility with either steric repulsion or with an
aligning/anti-aligning
interaction. Thus, it has been shown that at sufficiently high mean particle density, phase separation may occur in two-dimensional 
systems of repulsive finite-sized swimmers or self-propelled discs \cite{Buttinoni_13,Bialke13}. This result is explained by the 
self-trapping of 
colliding
swimmers, i.e., any two swimmers that collide and swim against each other remain in contact for a certain 
time span until their swimming directions have sufficiently changed. Recently, it was shown that a mixture of short-range aligning 
and long-range anti-aligning interactions between point-like active Brownian particles leads to a rich variety of density and velocity 
patterns in dry systems without memory \cite{Grossmann14} and with memory \cite{Aranson15}.


In contrast to previous studies, we demonstrate here that emergent collective dynamics in the form of persisting regular and irregular meso-scale density patterns can also be found in colonies of self-propelled particles that do not interact directly.
To this end, we consider active Brownian surfactant particles that move on the deformable surface of a thin liquid layer supported by a solid 
substrate.
The direction of swimming of each particle is assumed to have a non-zero vertical component, thus, leading to particles 
pushing against the film surface. Variations in the particle density give rise to large-scale film surface deformations that 
in turn induce flow in the layer of viscous fluid,
which drives even more
particles by advection
towards denser regions and also rotates
their swimming directions.
Particle diffusion and in particular the Marangoni effect act stabilizing. We
assume that the swimmers act as a surfactant, i.e., the local surface tension 
decreases
with swimmer 
concentration. This results in a soluto-Marangoni effect, i.e., Marangoni forces due to concentration gradients act at the free 
surface of the liquid film - they are a direct consequence of entropic contributions to the free energy of the free interface plus 
surfactant system \cite{ThAP2012pf}. In consequence, these Marangoni forces act stabilizing by driving the liquid away from 
areas with increased particle concentration.
Thus, in
this system the interaction between the particles is indirect and only occurs when the liquid layer dynamically deforms.

The first model system of swimmers on a liquid carrier layer was introduced in Ref.\,\cite{AlMi2009pre}. There it is assumed that the 
particles swim exclusively upwards at all times, and are not able to move along the film surface by self-propulsion. It is shown that 
the resulting excess pressure onto the film surface may cause a long-wave deformational instability of the film. The picture becomes 
more 
diverse,
when one allows for lateral 
active
motion of the swimmers as well \cite{PoTS2014pre}. Then,
a sufficiently large swimming velocity and 
a moderate rotational diffusion strength can suppress the long-wave instability due to the excess pressure discussed in Ref.~\cite{AlMi2009pre}.  In Ref.~\cite{PoTS2014pre} the linear stability results are confirmed by hybrid (multiscale) discrete-continuous 
numerical simulations, but fully nonlinear results obtained with the continuum model become available only now.

The paper is organized as follows: In Section~\ref{Sec1} we derive the coupled long-wave evolution equations for 
the space- and time-dependent full
swimmer density, 
which also includes swimmer orientation
(Smoluchowski equation), and the space- and time-dependent film profile (thin film equation). Next, we present a detailed 
linear-stability
analysis 
of the trivial steady state, i.e., of a homogeneous distribution of swimmers without preferred swimming direction (orientation) on the surface of a flat film. Different instability modes are discussed and located in a stability diagram 
spanned by rotational diffusion and self-propelling velocity. In Section~\ref{Sec2} we discuss various 
spatio-temporal
patterns that emerge in the nonlinear regime,
when the swimmers self-organize into persisting non-uniform structures.
In particular, we analyse stable standing and travelling density
waves accompanied by film modulation waves,
travelling waves that are modulated by large scale structures, and irregular patterns. 
We discuss the multistability of several of these states in 
a certain region
of the parameter space.

%
\section{Motion of active Brownian swimmers at slowly deforming interfaces}
\label{Sec1}
\begin{figure}[ht]
\centering
\includegraphics[width=0.5\textwidth]{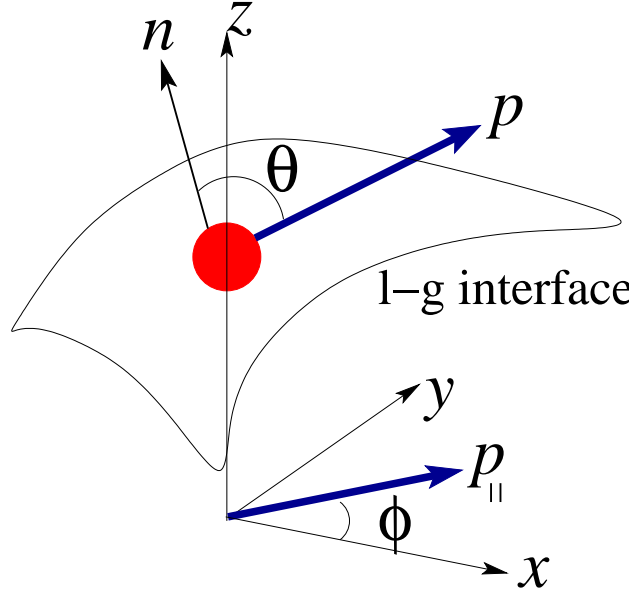}
\caption{(Color online) {Detail of the liquid-gas interface 
with height profile $h(x,y)$ 
and
with a single swimmer (filled circle). The instantaneous swimming direction is given by the vector ${\bm p}$ that makes an angle $\theta$ with the local normal ${\bm n}$. ${\bm p_\parallel}$ is the projection of ${\bm p}$ onto the $(x,y)$ plane. 
}
  \label{F1}}
\end{figure}
%

We consider a $10$-$100$ $\mu$m thin liquid film on a solid plate with a time-dependent film thickness profile $h(x,y,t)$.
The deformable liquid-gas interface (shortly called ``free surface'')
is covered by a colony of non-interacting active Brownian particles.
Besides
being microswimmers
the particles act as insoluble surfactants \cite{lubrication, Schwartz95,ThAP2012pf} as often found for small particles \cite{GaCS2012l}. In this way the particles are confined to move along the free surface and their density influences the interfacial energy of the free surface.
The lower bound for the average film thickness of $10~\mu$m is dictated by the 
typical size of
self-propelling mono-cellular 
organisms
such as {\it E.coli} or 
{\it the African trypanosome} \cite{Alizadehrad15}, 
%
or artificial swimmers such as 
phoretically
driven Janus particles 
(see, for example,
\cite{Howse07,C4SM02833C}). 
The typical size $R$ of such swimmers is of the order of $ R \approx 1~\mu$m although recently much smaller swimmers of $R\approx30$~nm have also been created \cite{LAMH2014nl}.
In what follows we assume a dilute limit, 
where
the average separation distance between the swimmers is much larger than their size. In this regime, 
direct two-particle interactions
as well as 
hydrodynamic interactions 
can be neglected.


We start by deriving the equations of motion for an active Brownian particle that moves 
along a two-dimensional time-dependent surface profile $h(x,y,t)$, as shown schematically in Fig.~\ref{F1}. The three-dimensional position vector of the particle is given by ${\bm r}=(x,y,h(x,y,t))$. The kinematic equation for the velocity reads 
\begin{eqnarray}
\label{vel}
{\dot{\bm r}}=(\dot{x},\dot{y},\partial_t h + \dot{x} \partial_x h +\dot{y}\partial_y h ).
\end{eqnarray}
Note that for fixed surface shape the velocity vector $\dot{\bm r}$ and the unit vector normal to the surface ${\bm n} = (-\partial_x h, -\partial_yh, 1) / 
\sqrt{1+(\partial_xh)^2+(\partial_y h)^2}$ are orthogonal
to each other:  
$\dot{\bm  r} \cdot {\bm n}=0$.  

In the overdamped limit the total velocity of the particle moving along the interface $h(x,y,t)$ is 
given by 
the superposition of the local tangential components of the self-propulsion velocity $v_0 {\bm p}_\tau$, the local tangential fluid velocity ${\bm u}_\tau$
of the film,
the tangential component of the gravity force ${\bm g}_\tau=-g({\bm e}_z)_\tau$, and 
thermal
noise ${\bm \eta}_\tau(t)$, 
which
results in diffusion along the free surface,
%
\begin{eqnarray}
\label{kinetic}
\dot{{\bm r}}=v_0 {\bm p}_\tau + {\bm u}_\tau + M\,m{\bm g}_\tau + {\bm \eta}_\tau(t) \, .
\end{eqnarray}
%
Here,
$M$ 
denotes
the mobility of the particle, $m$ 
its effective mass, which is reduced due to buoyancy effects for partly submerged particles, and the unit vector ${\bm p}$ indicates the direction of swimming. 
Thermal noise is 
characterized
by a Gaussian random variable with zero mean and
correlation function 
$\langle {\bm \eta}(t) {\bm \eta}(t^\prime)\rangle = 2Mk_B T \delta(t-t^\prime)  
{\bm 1}_2  $, where $k_B$ is the Boltzmann constant, $T$ is the absolute temperature, and 
${\bm 1}_2$ is a $2 \times 2$ unit matrix.
Note that the tangential component ${\bm a}_\tau$ of any vector ${\bm a}$ is given by 
\begin{eqnarray}
\label{tau}
{\bm a}_\tau = {\bm a}- ({\bm a}\cdot {\bm n}) {\bm n}.
\end{eqnarray}
%
%
%
Furthermore, the fluid velocity
${\bm u}$ at the 
free surface satisfies the standard kinematic boundary condition resulting from continuity \cite{lubrication}
\begin{eqnarray}
\partial_t h = -u_x \partial_x h -u_y \partial_yh  + u_z.
\end{eqnarray}

We consider particles swimming upward against gravity  and pushing against the film surface. 
This already creates some polar order with a preferred vertical orientation of the swimmer bodies
at the interface \cite{EncStark11}.  
%
%
Further reasons for such a polar order can be bottom-heaviness \cite{wolf13},
the  chemotactic response of bacteria swimming towards the surface in order to take up oxygen \cite{dombrowski04},
or any mechanism at the interface that aligns the particles along the vertical. In the following, we will not present a
full derivation of the orientational distribution at the interface. Instead, for the distribution against the surface normal 
we will assume that it always adjusts instantaneously compared to the slow dynamics of the film interface (see below).

In what follows, we only take into account the long-wave deformations of the film surface,
thus
$\epsilon = h_0/\lambda \ll 1$, with 
$\lambda$
the wavelength of the surface deformations 
and 
$h_0$
the average film thickness.
By noticing that the in-plane gradient $(\partial_x,\partial_y )$ is of order $\epsilon$, we obtain 
for the surface normal
${\bm n}=(0,0,1) + O(\epsilon)$ and for any vector ${\bm a}$ one has ${\bm a}_\tau=(a_x,a_y,0) + O(a_z\epsilon)+ O(a_x\epsilon)+O(a_y\epsilon)$, as it follows from Eq.\,(\ref{tau}). 
Then, the Langevin equation\ (\ref{kinetic}) for the interfacial particle position becomes
in 
leading order of $\epsilon$,
%
\begin{eqnarray}
\label{kinetic_xy}
 \dot{x}&=&v_0 {\bm p}_x + {\bm u}_x + \eta_x(t),\nonumber\\
 \dot{y}&=&v_0 {\bm p}_y + {\bm u}_y + \eta_y(t) \, .
\end{eqnarray}
Note that the tangential component of the gravity field vanishes in the long-wave limit, i.e. ${\bm g}_\tau=0$.

%

The instantaneous
orientation of swimmers is 
indicated
by the three-dimensional unit vector ${\bm p}$, as shown in Fig.\ref{F1}. For swimmers in the bulk of the fluid, the time-evolution 
of ${\bm p}$ is well known: it is determined by the rotation due to 
local fluid vorticity, alignment against 
some external field such as gravity,
and random rotation with the rate controlled by the rotational diffusivity $D_r$. However, for 
swimmers at a free surface, the rate of change $\dot{\bm p}$ may be significantly modified as compared to 
bulk swimmers depending on the nature of the interaction between the swimmers and the free film surface. 
For instance, a surface swimmer only partly submerged in the fluid and possibly with elongated body shape is easily 
rotated by local fluid vorticity within the film surface. However, the rotational rate of ${\bm p}$ against the interface normal
is possibly reduced as the interaction energies change with orientation of the swimmers at the free surface similar to anchoring effects for liquid crystals.

Here, we 
refrain
from deriving the exact evolution equation for the orientation vector ${\bm p}$ of partly submerged surface swimmers. Instead, 
we use the
argument from above
to decouple the vertical component ${\bm p}_\perp$ from the in-plane component ${\bm p}_\parallel$. Thus, we assume that 
the evolution of the film surface is slow and the equilibration of ${\bm p}$ to a 
stationary distribution $P_s(\theta)$ 
with respect to the vertical
is fast. 
In this case, the 
mean
vertical 
component of ${\bm p}$ 
is
given by
\begin{eqnarray}
\label{components}
\langle {\bm p}_\perp\rangle=\int_0^\pi P_s(\theta)\cos{\theta}\,
\sin{\theta}\,d\theta,
\end{eqnarray}
%
whereas the in-plane component ${\bm p}_\parallel$ can vary according to the in-plane dynamics
of ${\bm p}$, which couples to the temporal film evolution.
Note that $P_s(0)>P_s(\pi)$,
which implies
that the swimmers 
push
on average against the liquid-gas interface.

As a result, the
rotation of the in-plane component ${\bm p}_\parallel$ is 
described in terms of the polar angle $\phi$ (see Fig.\ref{F1})
\begin{eqnarray}
\label{rotation}
\dot{\phi} = \frac{1}{2}\Omega_z +\chi(t)  \, ,
\end{eqnarray}
where $\Omega_z= \partial_x u_y -\partial_y u_x$ is the vertical component of the local fluid vorticity and $\chi(t)$ is 
rotational noise with 
correlations
$\langle \chi(t)\chi(t^\prime)\rangle = 2D_r\delta(t-t^\prime)$. 
Furthermore, we introduce the mean in-plane velocity of an active particle, 
$v_\parallel = v_0  [1 - \langle {\bm p}_\perp \rangle^2]^{1/2}$ and
substitute 
$v_0 {\bm p}_\parallel$ in Eq.\,(\ref{kinetic_xy}) by $v_\parallel {\bm q}  =  v_\parallel (\cos{\phi},\sin{\phi})$. 
Then, the
Smoluchowski equation for the 
particle probability density
$\rho(x,y,\phi,t)$, equivalent to Eqs.\,(\ref{kinetic_xy}) and 
(\ref{rotation}), 
reads
\begin{eqnarray}
\label{smoluch}
\partial_t \rho+{\bm \nabla}\cdot {\bm J}_t + \partial_\phi J_{r}=0,
\end{eqnarray}
where ${\bm \nabla}=(\partial_x,\partial_y)$ and the 
respective 
 translational (${\bm J}_t$) and rotational ($J_r$) probability currents 
become 
 \cite{Gard,ps12,PoTS2014pre} 
%
\begin{eqnarray}
\label{currents}
 {\bm J}_t &=&(v_\parallel {\bm q}+ {\bm u}_\parallel)\,\rho - Mk_B T ({\bm \nabla} \rho ),\nonumber\\
 {J}_r &=& -D_r \partial_\phi \rho + \frac{1}{2}\Omega_z\rho.
\end{eqnarray}
%


The
swimmers and the 
liquid-gas interface couple to each other
through the local swimmer concentration $\rho(x,y,\phi,t)$ that acts twofold. 
First,
each swimmer exerts 
the force $\alpha=v_0 \langle p_\perp\rangle /M$ in the direction normal to the surface \cite{AlMi2009pre}.
So, the
total pushing force $f_n$ of the swimmers per unit area 
is proportional to the direction-averaged local concentration of swimmers, 
$\langle \rho \rangle (x,y,t) = \int_0^{2\pi} \rho(x,y,\phi,t)\,d\phi$, 
and becomes
\begin{eqnarray}
\label{push}
f_n(x,y,t) = \alpha \langle \rho\rangle(x,y,t) \, .
\end{eqnarray}
%
Secondly, the swimmers act as a surfactant
and change
the local surface tension. Assuming a relatively low concentration of swimmers, the surface tension $\sigma$ is known to decrease linearly with the local direction-averaged concentration $\langle \rho \rangle (x,y,t)$ \cite{ThAP2012pf},
\begin{eqnarray}
\label{marangoni}
\sigma= \sigma_0 - \Gamma \langle \rho \rangle \, ,
\end{eqnarray}
with the reference 
surface tension $\sigma_0$ and $\Gamma>0$.

Through Eqs.\ (\ref{push}) and (\ref{marangoni})
the Smoluchowski equation\
(\ref{smoluch}) and the thin film equation in the long-wave approximation \cite{lubrication} 
for the local film thickness $h(x,y,t)$
are coupled to each other \cite{PoTS2014pre,AlMi2009pre},
\begin{equation}
\label{thin_film}
\partial_t h+{\bm \nabla}\cdot\left( \frac{h^3}{3\mu}{\bm \nabla}\left[\sigma_0 \Delta h - \rho_l g h + \alpha\langle \rho \rangle \right]\right) -\Gamma {\bm \nabla}\cdot\left(\frac{h^2}{2\mu}{\bm \nabla}\langle \rho \rangle\right)=0,
\end{equation}
where $\rho_l$ is the density of the fluid and $\mu$ is its dynamic
viscosity.  The in-plane fluid velocity at the interface, ${\bm
  u}_\parallel=(u_x,u_y)$, is determined by the film profile
$h(x,y,t)$ \cite{lubrication},
\begin{eqnarray}
\label{surface_vel}
{\bm u}_\parallel&=&-\frac{\Gamma}{\mu} h{\bm \nabla}\langle \rho\rangle + \frac{h^2}{2\mu}{\bm \nabla}\left( \sigma_0\Delta h + \alpha\langle \rho \rangle \right),
\end{eqnarray}
%
and the
vertical component of the vorticity is obtained as $\Omega_z= \partial_x u_y -\partial_y u_x$
from Eq.\,(\ref{surface_vel}) \cite{note_omega}.
Both, ${\bm u}_\parallel$ and $\Omega_z$ enter the currents (\ref{currents}) that determine the Smoluchowski equation\ (\ref{smoluch}).
Note, that without swimming along the interface and rotational diffusion, one can integrate this equation over $\phi$ to recover the model 
in Ref.~\cite{AlMi2009pre} with purely upwards pushing swimmers. Switching off the active swiming motion altogether one recovers 
the classical long-wave model for a dilute insoluble surfactant on a liquid film that may be written in a gradient dynamics form \cite{ThAP2012pf}.

In what follows we focus on the instability induced by the pushing force generated by swimmers that swim predominantly upwards. To this end, we neglect the stabilising effect of the hydrostatic pressure $\rho_l g h_0$ as compared with the typical pushing force 
per unit area $\sim \rho_0v_0/M$.
Experimentally, such a regime can be achieved by using,
for example,
bacteria-covered 
water
films 
with a dense bacterial coverage.
In the dilute limit treated in this manuscript, one needs
conditions of microgravity.
To illustrate this example, we present some estimates.
The
maximal self-propulsion force of a unicellular bacterium is known to be 
of
the order of several $pN$ \cite{nature_force}.
The maximal surface density $\rho_0$ is estimated as $\rho_0\approx R^{-2}$, where $R\approx 1\mu$m is the typical size of the 
bacterial
body. The dilute limit corresponds to densities of at least one order of magnitude below $R^{-2}$. Consequently, the maximal pushing force per unit 
are
is estimated as $\rho_0 v_0/M \approx 10^{-1} N/{\rm m}^2$. On the other hand, $\rho_l g h_0 \approx g\,10^{-2} N/{\rm m}^2$ for a $10~\mu$m thick water film. Clearly, $\rho_l gh_0$ can be neglected 
against
$\rho_0 v_0 /M$ in case of 
$g \ll 10\,{\rm m}/s^2$.

The possibility to experimentally detect the thin-film instability due to the pushing force
$\alpha$
exerted by 
self-phoretic
particles is further strengthened by recent experiments with $\sim 30$~nm small Janus particles \cite{LAMH2014nl}. 
A much smaller particle size allows for larger surface particle densities and
may give
rise to larger excess pressure. In fact, the maximal density increases as $\sim R^{-2}$
for decreasing
particle size $R$.
However, it remains unclear how the pushing force 
$\alpha$
of a single 
self-phoretic
particle scales with its size. If the decrease of the pushing force with $R$ is slower than $\sim R^{2}$, the resulting 
excess pressure 
$\sim \rho_0 f$
exerted by the particles onto the liquid-gas interface can be several orders of magnitude larger than the value of 
$\approx 10^{-1} N/{\rm m}^2$ estimated
before in the dilute limit of a bacterial carpet.

For all what follows, we non-dimensionalise the evolution equations for film thickness $h(x,y,t)$ and swimmer
density $\rho(x,y,\phi,t)$ employing the scaling as in Ref.~\cite{PoTS2014pre}. 
Thus, we use $h_0$ as the vertical length scale, $h_0\sqrt{\sigma_0/\Gamma\rho_0}$ as the horizontal length scale, $\mu h_0 \sigma_0/(\Gamma^2 \rho_0^2)$ as the time scale, and the direction-averaged density of swimmers in the homogeneous state $\rho_0$ as the density scale. This gives the relevant parameters of our model:
the dimensionless in-plane self-propulsion velocity $V=v_\parallel\mu\sigma_0^{1/2}/(\Gamma\rho_0)^{3/2}$, the dimensionless in-plane rotational diffusivity $D=D_r h_0 \mu\sigma_0/(\Gamma\rho_0)^2$,
the translational surface diffusivity $d=k_B T M \mu/(h_0\rho_0\Gamma)$, and the excess pressure parameter  
$\beta=\alpha h_0 /\Gamma$. Furthermore, we introduce the effective in-plane diffusivity 
\begin{eqnarray}
\label{eff_diff}
D_{\rm eff} = \frac{V^2}{2D}+d \, .
\end{eqnarray}
Note that $D_{\rm eff}$ corresponds to the diffusion coefficient of a single self-propelled Brownian particle moving along 
a flat two-dimensional surface \cite{Palacci10,tenHagen11b}. 
%
%
In Appendix \ref{AppendixA} we summarize our non-dimensionalised dynamic equations.

%
\section{Linear stability of a flat film with homogeneously distributed swimmers}
\label{LinStab}
\subsection{General}
\label{LinStabGen}
%
%
%
%
We start by presenting more details of our stability analysis of the flat film as compared to our previous work 
\cite{PoTS2014pre} including an analytical treatment and a more thorough discussion of the occuring dispersion relations.
%
%
%
We linearise the non-dimensionalized Eqs.~(\ref{smoluch}) and (\ref{thin_film}) about the homogeneous isotropic steady state given by $h=1$, $\rho=1$, using the ansatz
\begin{eqnarray}
\label{subs}
h(x,y,t)=1+\delta h,~~\rho(x,y,\phi,t)= 1+\delta \rho,
\end{eqnarray}
where $\delta \rho, \delta h \ll 1$. The linearised
Smoluchowski equation\ (\ref{smoluch}) and the
thin film equation~(\ref{thin_film}) become, respectively,
\begin{eqnarray}
\label{smoluch_dimless}
\partial_t \delta \rho + {\bm \nabla }\cdot \left(V {\bm q} \delta \rho \right)+ \Delta \left[\left(\frac{1}{2}\beta -1\right)\langle \delta \rho\rangle+\frac{1}{2}\Delta (\delta h)\right]\frac{1}{2\pi} - d\Delta  \delta \rho - D \partial^2_\phi  \delta \rho=0. \\
\label{thin_film_lin}
\partial_t( \delta h)+\frac{1}{3}\left[\Delta^2 (\delta h) + \beta \Delta \langle \delta \rho \rangle  \right] -\frac{1}{2}\Delta \langle \delta \rho \rangle=0,
\end{eqnarray}
%
%
with  $\langle \delta \rho \rangle = \int_0^{2\pi} \delta \rho(x,y,\phi,t)\,d\phi$. The linearised surface velocity 
$(\delta u_x,\delta u_y)$ 
from Eq.\ (\ref{surface_vel})
reads
\begin{equation}
\label{vel_lin}
\delta {\bm u}_\parallel=-{\bm \nabla}\langle \delta \rho \rangle + \frac{1}{2}{\bm \nabla}\left[\Delta (\delta h)+\beta \langle \delta \rho \rangle\right] \, .
\end{equation}

Next, we follow \cite{PoTS2014pre} and Fourier transform the perturbations $\delta h$ and $\delta \rho$ by using a continuous Fourier transform in space and a discrete Fourier transform in the angle $\phi$.
%
Combining this with an exponential ansatz for the time evolution of the individual modes we have
\begin{eqnarray}
\label{expansion}
\delta
h({\bm r},t)&=& \int \hat{h}({\bm k})e^{\gamma({\bm k})t}e^{I{\bm k}{\bm r}}\,d{\bm k},\nonumber\\
\delta \rho({\bm r},\phi,t)&=&\lim_{N\rightarrow \infty} \frac{1}{2\pi}\sum_{n=-N}^{N} e^{I n \phi}\int W_n({\bm k})e^{\gamma({\bm k})t}e^{I{\bm k}{\bm r}}\,d{\bm k},
\end{eqnarray}
with the small dimensionless Fourier amplitudes $\hat{h}({\bm k})$ and $W_n({\bm k})$, the wave vector of the perturbation ${\bm k}=(k_x,k_y)$, and the real or complex growth rate $\gamma({\bm k})$.
Substituting the expansions 
from
Eqs.\,(\ref{expansion}) into the linearized Eqs.\,(\ref{smoluch},\ref{thin_film}), we obtain the eigenvalue problem 
\begin{equation}
\label{eig_lin}
\gamma({\bm k}) \mathbf{H} = {\mathcal{J}}({\bm k}) \mathbf{H},
\end{equation}
with the eigenvector $\mathbf{H}$
\begin{eqnarray}
\label{eig_vector}
{\mathbf H}({\bm k})=(\hat{h}, W_0, W_1, W_{-1}, W_2, W_{-2}, \dots),
\end{eqnarray}
and the Jacobi matrix $\mathcal{J}$, which corresponds to a banded matrix of the structure
\begin{eqnarray}
\label{Jacobi}
-{\mathcal{J}}({\bm k}) = \left(
\begin{array}{ll|llllll}
T_{11}, & T_{12}, & 0 & 0 & 0 & 0&0& \dots \\
T_{21}, & T_{22}, & V^{(-)}, & V^{(+)}, & 0 & 0&0&\dots \\
\hline
0 & V^{(+)}, & D +dk^2,& 0 & V^{(-)}, & 0, & 0&\dots \\
0 & V^{(-)}, & 0 & D+dk^2, &0 & V^{(+)}, & 0&\dots \\
0 & 0 &  V^{(+)}, & 0 & 2^2D+dk^2, & 0 & V^{(-)} & \dots \\
0 & 0 & 0 & V^{(-)}, & 0 & 2^2D+dk^2, &  0 & \dots \\
0 & 0 & 0 & 0& V^{(+)}, & 0 &3^2D+dk^2 &  \dots \\
\dots
\end{array}
\right),
\end{eqnarray}
where $k^2=k_x^2+k_y^2$, $V^{(+)}=\frac{Vk_y}{2}+ \frac{iVk_x}{2}$, $V^{(-)}=-\frac{Vk_y}{2}+ \frac{iVk_x}{2}$. The $(2\times 2)$ matrix ${\bm T}$ in the upper left corner of $\mathcal{J}$ 
is
given by
\begin{eqnarray}
\label{top_left}
{\bm T}({\bm k}) = \left(
\begin{array}{ll}
\frac{1}{3} k^4 , & \left(\frac{1}{2}-\frac{\beta}{3} \right) k^2 \\
\frac{1}{2} k^4, & \left(1 -\frac{1}{2}+d\right) k^2
\end{array}
\right).
\end{eqnarray}
%
Note that the matrix ${\bm T}$ coincides with the Jacobi matrix derived in Ref.\,\cite{AlMi2009pre} that encodes the linear stability in the special case of a flat film covered by autonomous purely upwards pushing motors that exert an excess pressure onto the liquid-gas interface.
In practice, we truncate the expansion in the angle $\phi$ and only take the first $N$ Fourier modes into account. Then, the Jacobi matrix $\mathcal{J}$ is a $(2N+2)\times (2N+2)$ matrix and the truncated eigenvector $\mathcal{H}=(\hat{h},W_0,W_1,W_{-1},\dots,W_N,W_{-N})$ is $(2N+2)$ dimensional. 
%
%

%
\begin{figure}[ht]
\centering
\includegraphics[width=0.95\textwidth]{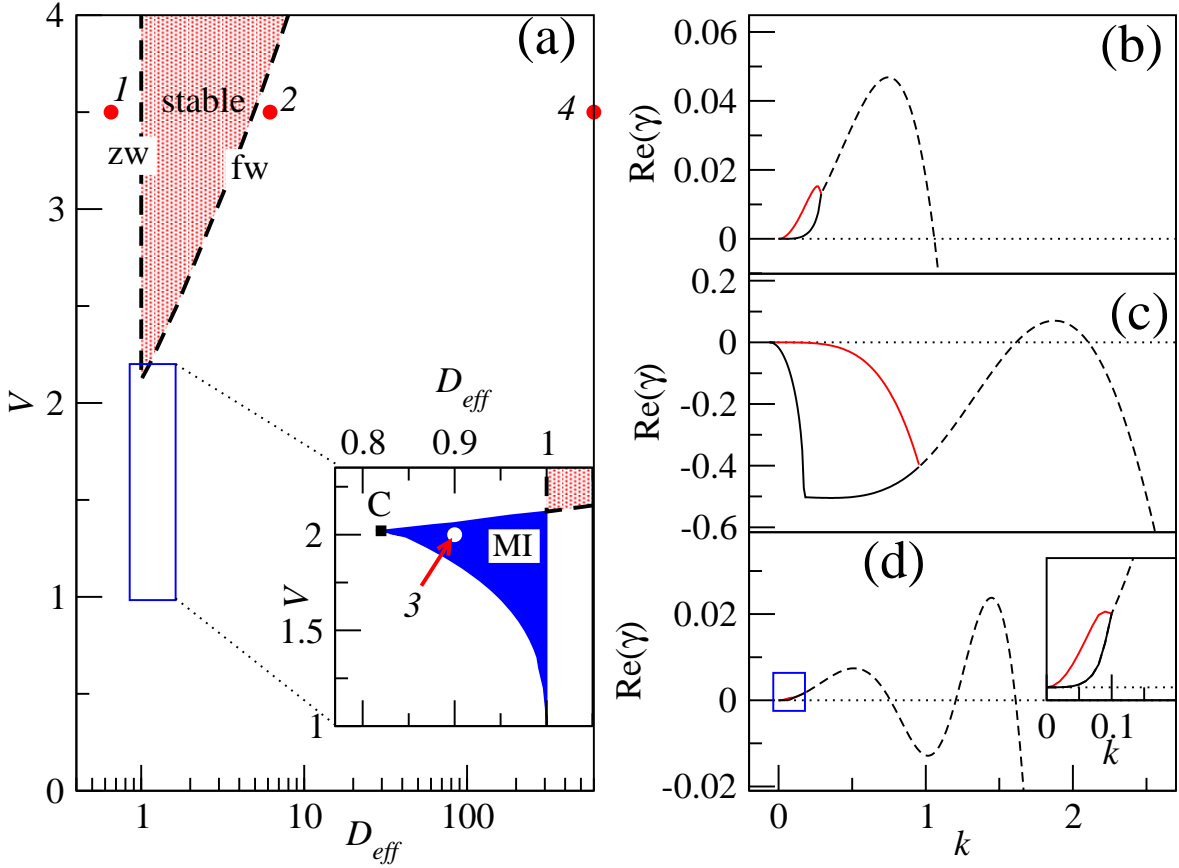}
\caption{(Color online) {(a) Stability diagram of a flat film with homogeneous distribution of particles for $\beta=4$ and $d=0.05$. 
In the stable lightly-shaded region Re$[\gamma(k)]<0$ for all values of the wave number $k$. The finite wave number instability sets in along the line marked by ``fw''. The zero wave number instability sets in along the vertical dashed line $D_{\rm eff}=1$, marked by ``zw''. 
The inset zooms 
into
the region marked by the rectangle in the main panel: In the strongly shaded area marked ``MI'', the mixed type fw /zw instability occurs. 
Panels (b-d) show Re$[\gamma(k)]$ of the two leading eigenvalues for parameters taken 
(b) at point $1$ ($V=3.5$, $D=10$, i.e., $D_{\rm eff}=0.66$), 
(c) at point $2$ ($V=3.5$, $D=1$, i.e., $D_{\rm eff}=6.175$), 
and (d) at point $3$ ($V=2$, $D=2.4$, i.e., $D_{\rm eff}=0.88$) in panel (a).  
Dashed and solid lines 
indicate
complex and real eigenvalues, respectively. The inset in (d) zooms 
into
the region marked by the rectangle. 
}
  \label{F2}}
\end{figure}
%

The stability diagram of the homogeneous isotropic 
state, as computed for
reduced excess pressure
$\beta=4$ and 
translational diffusivity
$d=0.05$ from Eqs.\,(\ref{eig_lin}) 
is shown in Fig.\,\ref{F1}(a)
in the parameter plane spanned by the reduced in-plane velocity $V$ and the effective diffusion constant
$D_{\rm eff} = V^2 / (2D) +d $ of the surfactants.
We numerically compute the eigenvalues of the truncated Jacobi matrix Eqs.\,(\ref{Jacobi}) for $N=10$ Fourier modes and then check the results by doubling the number of the Fourier modes to $N=20$.
Note that the choice $\beta=4$ and $d=0.05$ corresponds to a flat film that is unstable at zero in-plane velocity,
 $V=0$ as $\beta >\beta_c(d)=2(1+d)$, the critical value for the onset of the long-wave instability \cite{AlMi2009pre},
e.g., $\beta_c=2.1$ for $d=0.05$. In what follows, we characterize the system by the set of parameters $(V,D)$ and also indicate the respective value of $D_{\rm eff}$.

For the system with non-zero in-plane velocity ($V\not=0$), we have earlier reported the existence of two different instability modes \cite{PoTS2014pre}.
Namely, for sufficiently large $V$, there exists a wedge-shaped stability region, marked in Fig.\ref{F2}(a) by ``stable'' that separates regions where the two different instability modes occur. The wedge opens at $V_c\approx 2.05$ towards larger values of $V$, i.e., at any $V>V_c$, there exists a window in the effective diffusivity $D_{\rm eff}$ for which the flat homogeneously covered film is stable (note that $V_c$ depends on $\beta$ and $d$).

By crossing the two borders of the linearly stable region, the system changes stability via two distinct instability modes. The first mode 
corresponds to an oscillatory instability with a finite wave number at onset, i.e., a travelling wave instability. In this case, for parameters directly on the stability threshold, the leading eigenvalue $\gamma(k)$ of the Jacobi matrix 
from Eq.\ (\ref{Jacobi})
has a negative real part for all values of the wave number $k$, except for the
critical 
wave number
$k_c\not=0$, where $\gamma(k)$ has the form $\gamma(k_c) = \pm iW_c$ with some non-zero frequency $W_c$.
We will refer to this instability mode as the finite wave number instability (fw).
%
The second mode ($zw$)
corresponds to an instability with a zero wave number at onset.  This mode is characterized in detail in the next section.
\subsection{Zero wave number instability: analytic results}
%
%
%
%
In the following we present an approximate analytic expression for the zero wave number instability. We start by
introducing
the Fourier transformed fields $\hat{\delta h}({\bm k},t)$ and $\hat{\rho}({\bm k},\phi,t)$, according to $\delta h({\bm r})=\int e^{i{\bm k}\cdot {\bm r}}\hat{\delta h}({\bm k},t)\,d{\bm k}$ and $\rho(\bm r)=\int e^{i{\bm k}\cdot {\bm r}}\hat{\rho}({\bm k},\phi,t)\,d{\bm k}$, 
into 
Eq.\,(\ref{smoluch_dimless}) and obtain
%
\begin{eqnarray}
\label{smoluch_FT}
\partial_t \hat{\rho} + V\left( ik_x \cos{\phi},ik_y\sin{\phi}\right)\hat{\rho}+ \frac{1}{2\pi}\left(1-\frac{\beta}{2} \right)k^2\hat{\langle \rho\rangle}+\frac{1}{4\pi} k^4 (\hat{\delta h}) +d\,k^2 \hat{\rho} - D \partial^2_\phi \hat{\rho}=0.
\end{eqnarray}
%
%
Close to the threshold of the zero wave number instability, the amplitudes of all modes with the wave number $k\not= 0$ rapidly decay with time.  Therefore, in the limit $k\rightarrow 0$, in Eq.\,(\ref{smoluch_FT}) one may neglect the terms of orders $k^2$ and $k^4$ as compared to the ones of order $k^0$ and $k$. In consequence, the density and film height equations decouple. 
In fact, to this order
the density equation 
describes a single self-propelled particle with 
rotational diffusivity $D$ and self-propulsion velocity $V$
but with neglected translational diffusivity.

To proceed further, we note that on length scales larger than the persistence length of an active particle,  $V/ D$, 
the dynamics becomes purely diffusive. To arrive at this result, one performs a multipole expansion of 
$\hat{\rho}({\bm k},\phi,t)$ in the angle $\phi$ using only the monopol $\hat{\langle \rho\rangle}$ and the dipole moment
\cite{Golestanian12,Pohl14}. 
%
%
The latter can be elimated in the dynamic equation for $\hat{\langle \rho\rangle}$ and from Eq.\ (\ref{smoluch_FT})
one arrives at
%
%
%
%
\begin{eqnarray}
\label{Smol_3}
\partial_t \hat{\langle \rho\rangle}  + \left(1-\frac{\beta}{2}+D_{\rm eff}\right) k^2 \hat{\langle \rho\rangle}+\frac{1}{2}k^4 \hat{\delta h}=0,
\end{eqnarray}
which is coupled to the linearised thin film equation in Fourier space
\begin{eqnarray}
\label{thin_film_1}
\partial_t( \hat{\delta h})+\frac{1}{3}\left[k^4 \hat{\delta h} -\beta k^2 \hat{\langle \rho\rangle}  \right] + \frac{1}{2}k^2\hat{\langle \rho \rangle}=0.
\end{eqnarray}
Here, $D_{\rm eff} = V^2/(2D) + d$ is the effective diffusion constant of an active particle introduced earlier in 
Eq.\ (\ref{eff_diff}). The additional term results from the activity of the particle. Equations
(\ref{Smol_3}) and (\ref{thin_film_1}) are identical to the linearised evolution equations found for the concentration field of 
the purely upwards swimming ($V=0$)  surfactant particles coupled to the thin film equation, as studied in 
Ref.\cite{AlMi2009pre}. In fact, the results of the linear stability analysis of Ref. \cite{AlMi2009pre} can be translated to the system of equations~(\ref{Smol_3}) and (\ref{thin_film_1}) by setting the translational diffusivity of the purely upwards swimming particles to be equal to $D_{\rm eff}$.

The two eigenvalues $\gamma(k)$ resulting when introducing an exponential ansatz for the time dependence into  Eqs.\,(\ref{Smol_3}) and (\ref{thin_film_1}) are determined analytically (cf.\ Ref.~\cite{AlMi2009pre}). The Jacobi matrix ${\bm J}(k)$ of the linearised  Eqs.\,(\ref{Smol_3}) and (\ref{thin_film_1}) is given by
\begin{eqnarray}
{\bm J}=-k^2\,\left(
\begin{array}{cc}
\frac{k^2}{3}\qquad & \frac{1}{2}-\frac{\beta}{3} \\
\frac{k^2}{2}\qquad & \eta
\end{array}
\right),
\end{eqnarray}
where we introduced $\eta=1-\beta/2+D_{\rm eff}$.
%
The two eigenvalues $\gamma_{1,2}$ are
\begin{eqnarray}
\label{A1_gamma}
\gamma_{1,2}=\frac{1}{2}\left({\rm tr}({\bm J})\pm \sqrt{{\rm tr}({\bm J})^2-4{\rm det}({\bm J})}\right),
\end{eqnarray} 
with ${\rm tr}({\bm J})=-k^2[k^2/3+\eta]$ and ${\rm det}({\bm J})=k^6(1/12+D_{\rm eff}/3)$.

\begin{figure} 
\centering
\includegraphics[width=0.95\textwidth]{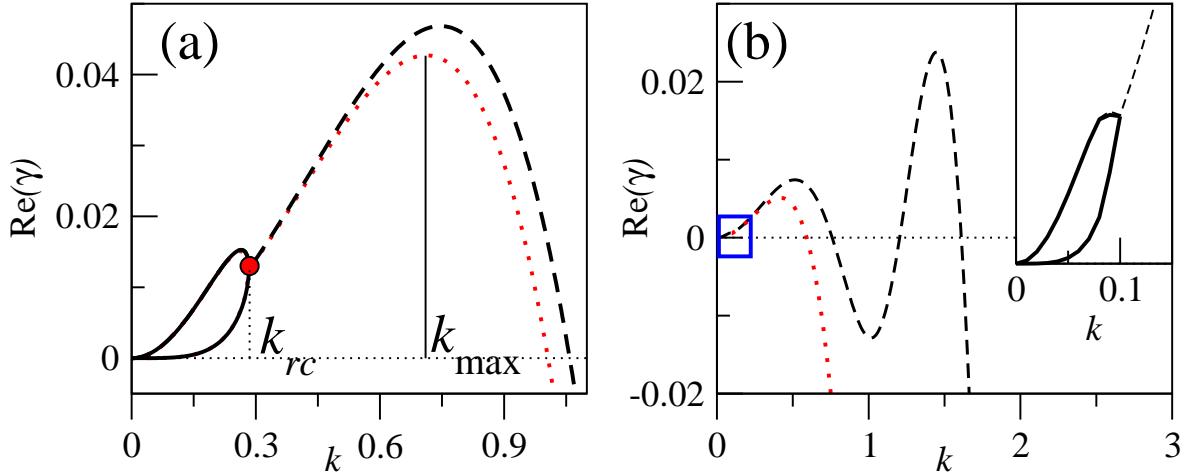}
\caption{(Color online)
Comparison of the numerically computed two leading eigenvalues (solid and dashed lines represent real and complex eigenvalues, respectively) and the analytic prediction 
of
Eq.\,(\ref{A1_gamma}) (dotted lines). (a) $V=3.5$, $D_{\rm eff}=0.66$, (b) $V=2$, $D_{\rm eff}=0.88$. The inset zooms 
into
the region marked by the rectangle,
where the
analytic and numerical results are indistinguishable
from each other.
In the inset we only show 
the analytically
computed eigenvalues.
}
\label{F3}
\end{figure}

In Fig.\,\ref{F3}(a,b) we compare the analytic eigenvalues given by Eq.\,(\ref{A1_gamma}) with the two leading eigenvalues of the original non-reduced system, computed numerically, as described in Section\,\ref{LinStabGen}. Solid 
and dashed
lines correspond to the numerically computed real and complex eigenvalues, respectively, dotted lines correspond to Eq.\,(\ref{A1_gamma}). Fig.\,\ref{F3}(a) is obtained for the parameters corresponding to point $1$ in Fig.\,\ref{F2}(a), i.e., $V=3.5$ and $D_{\rm eff}=0.66$, and 
Fig.\,\ref{F3}(b) corresponds to point $3$ in Fig.\,\ref{F2}(b), i.e., $V=2$ and $D_{\rm eff}=0.88$. In both cases, 
the agreement is excellent up to $k\approx 0.4$.

%
%
Next, we use Eq.\,(\ref{A1_gamma}) in order to classify the zero wave number instability that sets in along the dashed vertical line marked by ``zw" in Fig. \ref{F2}(a). 
Analysing
the eigenvalues in Eq.\,(\ref{A1_gamma}) shows that the real part of the leading eigenvalue changes its sign at $\eta=0$, or, equivalently at $D_{\rm eff}=\beta/2-1$.  Thus, for the value of $\beta=4$ used here, we obtain $D_{\rm eff}=1$, in agreement with Fig.\ref{F2}(a). We emphasize that the above analytic results can only be applied in the limit of $k\ll 1$, where the approximation~(\ref{Smol_3}) applies. 

We
find that in the unstable region, i.e., for $\eta<0$, the fastest growing wave number $k_{\rm max}$, 
indicated in Fig.\ \ref{F3}(a), always corresponds to a pair of two complex conjugate eigenvalues
[dashed line in Fig.\ \ref{F3}(a)].
By locating the maximum of 
${\rm tr}[J(k)]$, we 
%
determine $k_{\rm max}$ and its complex growth rate,
\begin{eqnarray}
\label{kmax}
k_{\rm max}=\sqrt{\frac{3}{2}\mid \eta\mid} ,\,\,\,{\rm Re}[\gamma(k_{\rm max})]=\frac{3}{8}\eta^2,\,\,\,{\rm Im}[\gamma(k_{\rm max})]=\pm\frac{\mid\eta \mid}{2}\sqrt{\frac{27}{2}\mid \eta\mid\left(\frac{1}{12}+\frac{D_{\rm eff}}{3}\right)-\frac{9}{16}\eta^2}.
\end{eqnarray}
For $k<k_{\rm rc}<k_{\rm max}$, the two leading
eigenvalues are real,
as 
indicated by
the two solid lines
in Fig.\ \ref{F3}(a).
%
The wave number $k_{\rm rc}$ is
\begin{eqnarray}
\label{A1_kc}
k_\mathrm{rc}= \sqrt{\frac{3\eta^2}{1+4D_{\rm eff}-2\eta}}\approx \sqrt{\frac{3}{1+4D_{\rm eff}}} \mid \eta\mid .
\end{eqnarray}
%
%
%
This result implies that the character of the zero wave number instability is peculiar: Directly at onset ($\eta=0$) the 
leading two eigenvalues are real, however, already arbitrarily close above onset ($\eta<0$) the band of unstable wavenumbers contains a region of real modes (close to and including $k=0$) and a region of complex modes (always including the fastest growing mode). This behaviour is related to 
the existence of two conserved fields, $h$ and $<\rho>$, that forces two real modes with growh rate zero at $k=0$. In consequence, the fastest
growing wave number $k_{\rm max}$ tends to zero when approaching the stability threshold from above. Here, we call this scenario 
 a zero-wave number instability.

\subsection{Mixed instability}
%
The analytic results obtained in the previous section give for the zero wave number instability the threshold $D_{\rm eff}=1$ 
for $\beta=4$. This perfectly coincides with the numerically computed threshold as indicated by the left thick dashed line 
in Fig.\ref{F2}(a). However, these results also remain valid around
$D_{\rm eff}=1$  for small values of $V<V_c$ deep in the unstable region. This implies
that regardless of the value of $V$, the sign of the dispersion curve 
${\rm Re}[\gamma(k)]$
at very small wave numbers $k\approx 0$ changes from negative to positive as $D_{\rm eff}$ is decreased past the critical value $D_{\rm eff}=1$ (at $\beta=4$).
However, there will always be an instability at a non-zero wave number, as we explain now.

We observe
a mixed-instability region, marked by "MI'' and heavily shaded in the inset of Fig.\,\ref{F2}(a),
where the system can be described as being unstable with respect to a mixed finite- and zero-wavelength instability. In this region, there exist two bands of unstable wave numbers: one with $k\in [0,k_1]$ and another one with $k\in[k_2,k_3]$, with $k_3>k_2 > k_1>0$.  
 The two leading eigenvalues that correspond to the pure zero wave number, the pure finite wave number, and the mixed instabilities, are shown in Figs.\,\ref{F2}(b), (c), and (d), respectively. The 
respective 
parameters 
are; point 1:
$V=3.5,\,D=10$, i.e., $D_{\rm eff}=0.66$,
point 2:
$V=3.5,\,D=1$, i.e., $D_{\rm eff}=6.175$ in the main panel, 
and point 3:
$V=2,\,D=2.4$, i.e. $D_{\rm eff}=0.88$ in the inset of Fig.\,\ref{F2}(a). 
Dashed 
and solid lines correspond,
respectively,
to complex and real eigenvalues.


When considering the type of
dispersion curves, the transition from the finite wave number instability to the zero wave number instability
can 
follow
two different scenarios.
We identify them by keeping $V$ constant and gradually decrease $D_{\rm eff}$.
In the first scenario
for $V \gtrsim V_c = 2.05$,
the maximum of 
${\rm Re}[\gamma((k)]$
in Fig.~\ref{F2}(c) first becomes negative, i.e., the finite wave number instability is stabilised
while crossing the "fw" line in the stability diagram.
Subsequently, the system crosses the "zw'' line and becomes unstable  w.r.t.\ the 
zero
wave number mode. 
In the second scenario
for $V \lesssim V_c=2.05$,
the system first crosses the line $D_{\rm eff}=1$ and enters the mixed instability region. Then the dispersion curve 
as in Fig.\ \ref{F2}(d) gradually transforms into the dispersion curve of the zero wave number instability,
while leaving the region MI. We also remark that the mixed instability region stretches from $V=V_c$ down $V=0$. However, its horizontal width is negligibly small for $V<1$.

\begin{figure}
\centering
\includegraphics[width=0.95\textwidth]{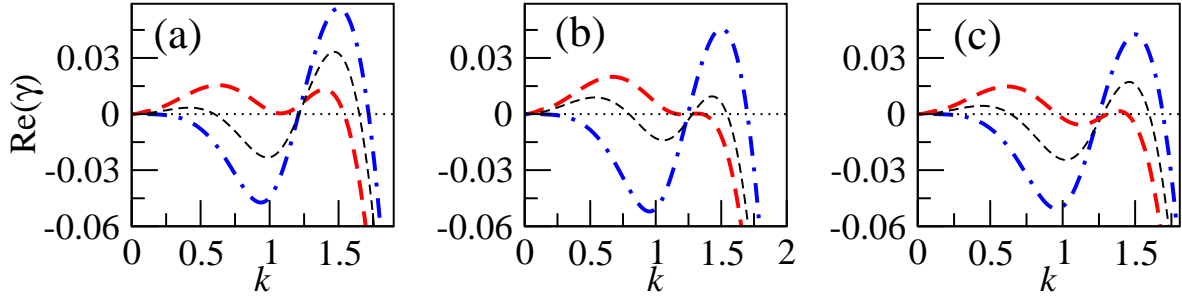}
\caption{(Color online)
Transition from the finite wave number to the zero wave number instability region 
while passing
through the mixed instability at fixed $V$ by increasing $D$ (decreasing $D_{\rm eff}$). In each panel $V$ is fixed and three different values of $D$ are chosen. The thick 
dotted-dashed (blue) lines correspond to $D_{\rm eff}=1$, the thick dashed (red) lines represent the critical dispersion curve 
on the border of the MI region in Fig.\ \ref{F2} at $D_{\rm eff}<1$,
and the thin dashed (black) line is for a value of $D_{\rm eff}$ in between.
The transitions are shown along the lines: (a) $V=2$, (b) $V=2.0275$ corresponding to the cusp point C in the inset of Fig.\ref{F2}(a)], 
and (c) $V=2.03$.
  \label{F4}}
\end{figure}

The second scenario is visualised in Fig.\ \ref{F4}. In Fig.\ \ref{F4}(a) we fix $V=2$ and plot ${\rm Re}[\gamma((k)]$ for three different $D_{\rm eff}$: $D_{\rm eff}=1$ (thick dotted-dashed blue line), $D_{\rm eff}=0.92$ (dashed black line) and $D_{\rm eff}=0.83$ (thick dashed red line).  In this case, the transition from the mixed instability to the zero wave number instability occurs through the elevation of the local minimum of the dispersion curve, so that a single band of unstable wave numbers occurs starting at $k=0$.

In Fig.\ \ref{F4}(b) we fix $V=2.0275$ what corresponds to the level of the cusp point marked by "C'' in the inset of Fig.\ \ref{F2}(a). We show ${\rm Re}[\gamma((k)]$ for $D_{\rm eff}=1$ (thick dotted-dashed blue line), $D_{\rm eff}=0.87$ (dashed black line) and $D_{\rm eff}=0.81$ (thick dashed red line).  In this case, the transition from the mixed instability to the zero wave number instability occurs through the simultaneous elevation of the local minimum and the depression of the local maximum of the dispersion curve. In Fig.\ref{F4}(c) we fix $V=2.03$ and plot ${\rm Re}[\gamma((k)]$ for $D_{\rm eff}=1$ (thick dotted-dashed blue line), $D_{\rm eff}=0.9$ (dashed black line) and $D_{\rm eff}=0.84$ (thick dashed red line).  In this case, the transition from the mixed instability to the zero wave number instability occurs through the depression of the local maximum of the dispersion curve. All dispersion curves in Fig.\ \ref{F4} correspond to real eigenvalues only for a very narrow band of the wave numbers $k \approx 0$ (details are not shown here).

\section{Nonlinear evolution}
\label{Sec2}
\subsection{Numerical approach and solution measures}
In this section we address the system of nonlinear evolution equations for film hight $h$ and probability density $\rho$,
which we give in non-dimensional form in Eqs.\ (\ref{nondim}) and (\ref{eq.several})
in appendix\ \ref{AppendixA}. In order to solve them
numerically, we
discretize
both
the film thickness $h(x,y,t)$ 
and the density $\rho(x,y,\phi,t)$ in 
a square box 
for the spatial coordinates with
$(x\in [-L/2,L/2]) \times (y \in [-L/2,L/2])$ 
and in the interval $\phi \in [0,2\pi]$ for the orientation angle always using
periodic boundaries.
We use $N=100$ or $N=128$ mesh points for each spatial direction
to discretise
space and $20$ Fourier modes for the decomposition of the $\phi$-dependence of the  density. We adopt a semi-implicit pseudo-spectral method for the time integration, as outlined in the Appendix and verify 
some of our
results by using a fully explicit Euler scheme with the time step of the order of $\Delta t =10^{-4}\dots 10^{-5}$.

In order to quantify the spatio-temporal patterns in film hight and density,
we introduce three global measures: the 
mode type $M$
with
%
\begin{eqnarray}
\label{int_mode}
M=L^{-2}\int\int (h(x,y,t)-1)(\langle \rho\rangle(x,y,t)-1)\,dxdy,
\end{eqnarray}
characterizes if spatial modulations of the film surface $h(x,y,t)$ and the average density $\langle \rho\rangle(x,y,t)$
are 
predominantly
in-phase 
($M> 0$) or 
predominantly in anti-phase
($M<0$);
%
%
the space-averaged flux of the fluid $\bar {\bm J}_h$ determined by 
%
\begin{eqnarray}
\label{int_flux}
\bar {\bm J}_h &=& L^{-2}\int\int \left(\frac{h^3}{3}{\bm \nabla}\left[\Delta h + \beta \langle \rho \rangle \right] -\frac{1}{2} \left(h^2 {\bm \nabla \langle \rho\rangle}\right)\right)\,dxdy
\end{eqnarray}
allows us to distinguish between standing waves that correspond to $\bar {\bm J}_h=0$ and travelling or modulated 
waves that are characterized by a non-zero fluid flux; and finally
the space-averaged translational flux of the swimmers,
\begin{eqnarray}
\label{int_Jt}
\bar {\bm J}_t &=& L^{-2}\int\int \left(V\langle \rho {\bm q}\rangle +{\bm U}_\parallel \langle \rho \rangle  \right)\,dxdy,
\end{eqnarray}
indicates
global surface motion of 
the
swimmers.
%

%

In the following we indicate the richness in the dynamics of our system by giving examples of evolving patterns for 
specific parameter sets located in the stability diagram of Fig.\ \ref{F2}(a). Mapping out a full state diagram is beyond the
scope of this article.

%
\subsection{Regular standing wave pattern}
\label{subsec.standing}



We first study the
parameters $V=3.5$ and $D=1$ ($D_{\rm eff}=6.175$) 
in the unstable region of the phase diagram Fig.\,\ref{F2}(a), close to the finite wave number instability threshold [point $2$ in 
Fig.\,\ref{F2}(a)]. The corresponding dispersion curve is shown in Fig.\,\ref{F2}(c). The system size $L=20$ is set to be several 
times larger than the fastest growing wave length equal to $2\pi/k_{\rm max}=3.43$, with $k_{\rm max}$ denoting 
the wave number corresponding to the maximum of the dispersion curve.
%
By numerically integrating 
Eqs.\ (\ref{nondim}),
we study the temporal evolution of the system from the homogeneous isotropic steady state $h=1$ and $\rho=1/(2\pi)$, i.e., the trivial state. The initial conditions are given by $h=1+\delta h(x,y)$ and $\rho=1/(2\pi)+\delta \rho(x,y,\phi)$, where the small amplitude random perturbations $\delta h(x,y)$ and $\delta \rho(x,y,\phi)$ represent two independent sources of 
white noise.

\begin{figure} 
\centering
\includegraphics[width=0.95\textwidth]{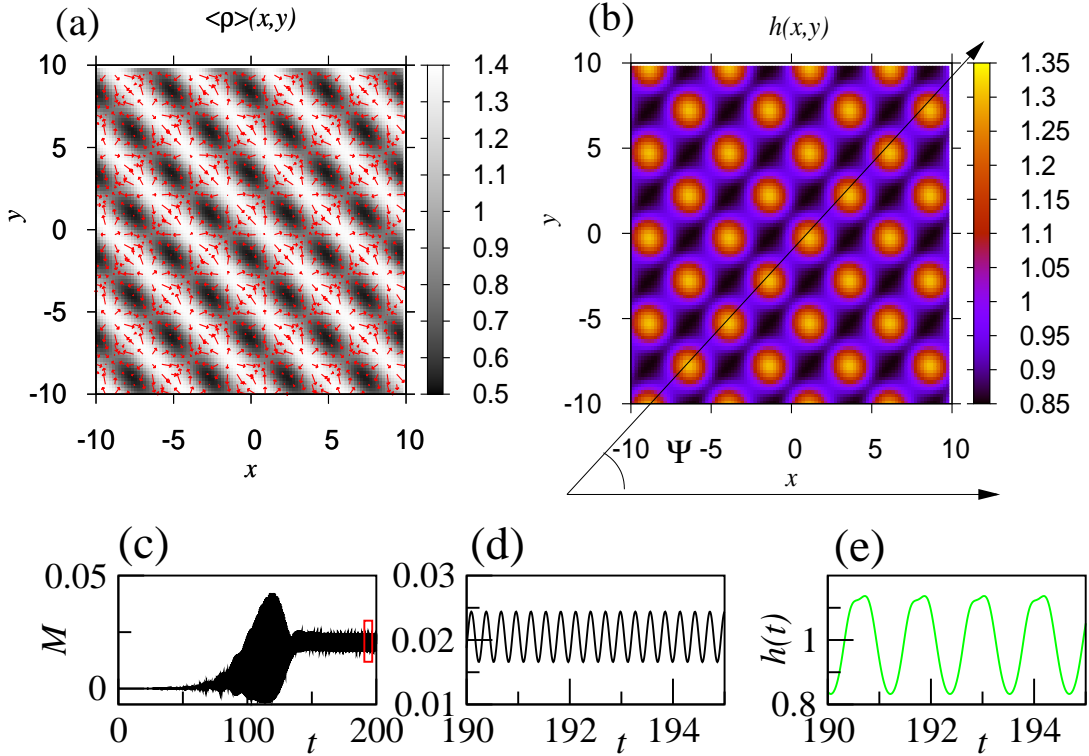}
\includegraphics[width=0.85\textwidth]{V3dot5_D1_new.eps}
\caption{(Color online)
Standing
wave pattern, obtained for $V=3.5$, $D=1$, $d=0.05$,
i.e., $D_{\rm eff}=6.175$
[the corresponding dispersion curve is shown in Fig.\,\ref{F2}(c)].
The snapshots are taken
at $t=190$: (a) average density $\langle \rho\rangle(x,y)$ (grey scale map) with average orientation field (red arrows) and (b) the film thickness $h(x,y)$. $\Psi$ is the 
angle between the main lattice direction
and the horizontal axis. 
Plotted versus time are:
(c) The mode type $M$ from Eq.\,(\ref{int_mode}), (d) zoom of the region in (c) marked by 
the red
rectangle, 
and (e) local film thickness $h(t)$ at a randomly chosen point on the surface. 
\label{F5}}
\end{figure}
\begin{figure}[ht]
\centering
\includegraphics[width=0.98\textwidth]{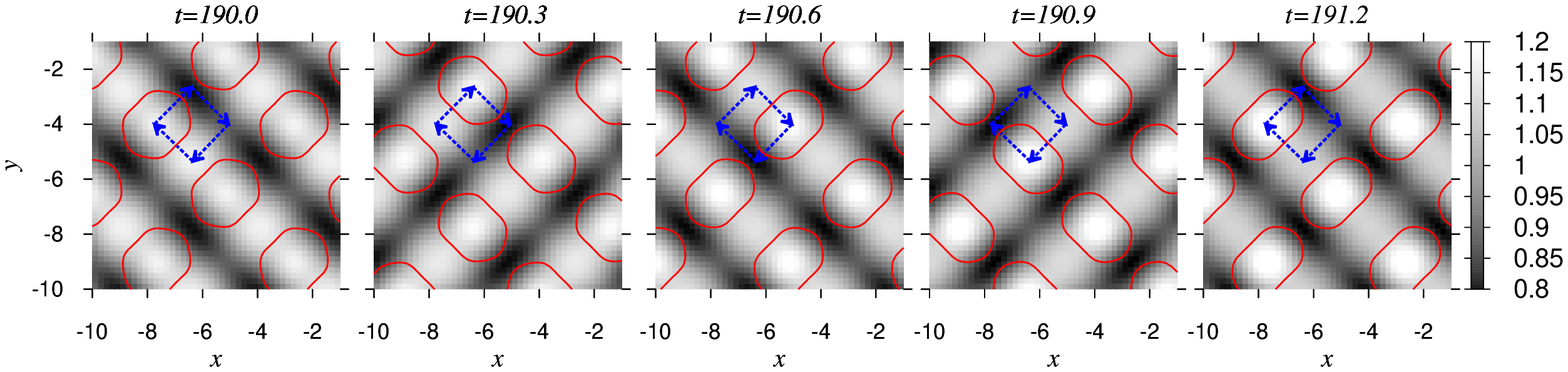}
\caption{(Color online) Standing 
stripe patterns with periodically changing directions for the same
parameters as in Fig.\,\ref{F5}. Shown is a small portion of the square domain in Fig.\,\ref{F5} over one oscillation cycle with 
period $T=1.15$.  Five snapshots of the film thickness $h(x,y,t)$ (in grey scale map) are 
plotted together with
a contour plot of the average density $\langle \rho\rangle$ at the level of $\langle \rho\rangle=1$. Arrows indicated the 
shift vector
of the pattern
during the four phases of one oscillation cycle.
}
\label{F6}
\end{figure}

We found that after a transient phase of an approximate duration of $100\dots 200$ time units, the system settles onto a stable time-periodic state. 
It
can be characterised as a regular standing wave,
where stripe patterns change periodically between the two diagonal directions as \href{https://www.dropbox.com/s/v4wm4l6a83dniee/V3d5_Deff_6d175.mp4?dl=0}{\color{red}{Movie 1}} shows.
The mean fluid flux is zero, $\bar J_h=0$.
Figure.\,\ref{F5}(a) shows a snapshot of the average density $\langle \rho\rangle(x,y)$ in grey scale map together with the average 
orientation field $ \langle {\bm q} \rangle = (\langle \cos{\phi} \rangle,\langle \sin{\phi} \rangle)$ shown by red arrows at 
time $t=190$. The corresponding snapshot of the film thickness $h(x,y)$ is shown in Fig.\,\ref{F5}(b). The patterns in 
Figs.\,\ref{F5}(a) 
and
(b) are highly dynamic, with the shape of the surfaces $h$ and $\langle \rho\rangle$ changing periodically in time, 
visualized in \href{https://www.dropbox.com/s/v4wm4l6a83dniee/V3d5_Deff_6d175.mp4?dl=0}{\color{red}{Movie 1}} and 
described in Fig.\ref{F6}.
Fig.\ \ref{F5}(b) indicates the moment in time when the orientiation of the stripes in the film profile changes from one diagonal to
the other. Maxima in the film height $h(x,y,t)$ occur, which
are arranged in a perfect square lattice, tilted by the angle $\Psi \approx \pi/4$ w.r.t. the $x$-axis.

The time evolution of the mode type $M$ starting from the 
initially 
homogeneous 
state is 
plotted
in Fig.~\ref{F5}(c). After $150$ time units $M$ starts to oscillate periodically about the average of $\overline{M}=0.02$, as indicated in Fig.~\ref{F5}(d), where the zoom of the region marked by the rectangle in Fig.~\ref{F5}(c) is shown.  As $M>0$, the oscillations of the film thickness and the averaged density are in-phase. 
The temporal period $T$ of the standing wave can be determined by observing the oscillations of the film thickness $h(t)$ at a randomly chosen point, as 
Fig.\,\ref{F5}(e)
demonstrates.
Thus $h(t)$ oscillates about the average film thickness $h=1$ as a perfect periodic function with the period $T=1.15$.

Remarkably, the temporal oscillations of $M$ in Fig.~\ref{F5}(d) 
are four times faster than the oscillations in film thickness.
This is due to the fact that a complete period of the standing wave consists of four phases,
where the same spatial pattern reappears four times, each time shifted along one side of a square and rotated by $90^{\circ}$. Because $M$ is invariant under rotation and translation of the pattern,
the oscillation period of $M$ is four times smaller then the overall period of the standing wave. The four shifted and rotated patterns are clearly seen in \href{https://www.dropbox.com/s/v4wm4l6a83dniee/V3d5_Deff_6d175.mp4?dl=0}{\color{red}{Movie 1}} when
concentrating on the cubic lattice formed by the maxima in the height profile. In Fig.~\ref{F6} we illustrate the four phases
by snapshots of the transient stripe patterns.
During the first phase, the maxima of the average density $\langle \rho \rangle$ shift along a straight line by a distance $l_w/2$ equal to half the spatial period $l_w$.
The maxima of the film thickness follow the same path. In each subsequent phase, the shift occurs along the direction that is orthogonal to the previous shift. 
After completing all four phases, the 
maxima of $\langle \rho \rangle$ and also of $h$ will have traveled along the sides of a square with the side length $l_w/2$ and have
returned to the initial position. In between the square patterns, formed by the maxima in density, the film thickness assumes patterns of parallel ridges that during each quarter of the cycle decay into the square pattern,
formed by the maxima in $h$,
and reappear rotated by $\pi/2$. One may say that during one cycle the pattern oscillates through several accessible 
patterns that are well known solutions for pattern forming systems on a square. In particular, they are known to occur as 
(stable or unstable) steady states in thin film equations that describe 'passive' liquid layers, ridges and drops on  
homogeneous solid substrates \cite{BeTh2010sjads}.

%
%
%

The spatial period of the standing wave, $l_w$, can be determined in real space by measuring the distance 
between two nearest maxima (minima) of the
height profile $h(x,y)$ taken at an arbitrary moment of time. 
The maximal error in this procedure is of the order of $\sqrt{2}L/N$
%
where $N$ is the number of discretization points  along the $x$ and $y$ axis
and the factor $\sqrt{2}$ reflects that the wave is directed along the diagonal of the domain.
We obtain $l_w=3.5\pm 0.14$ for the square patterns in Fig.\ \ref{F5}
using $L=20$ and $N=100$ to estimate the error.


Due to the periodic boundary conditions the
measured $l_w$ of the patterns in Fig.\ \ref{F5} is slightly different from the fastest growing wave length found from the dispersion
curve in Fig.\,\ref{F2}(b) as $l_{\mathrm{max}}=2\pi/k_{\mathrm{max}}=3.43$. This difference is explained as follows.
In order to fulfill periodic boundary conditions in a square domain of size $L$, the periods of a wave projected, respectively,
on the $x$ and $y$ axis are $L/k$ and $L/m$, where $m$ and $k$ are some integers.
This restricts the possible rotation angles $\Psi$ of a periodic pattern relative
to the $x$ axis (see Fig.\ \ref{F5}). They have to satisfy
\begin{eqnarray}
\label{Psi}
\cos{\Psi} = \frac{m}{\sqrt{k^2+m^2}},\,\,\,\sin{\Psi}=\frac{k}{\sqrt{k^2+m^2}}
\end{eqnarray}
%
and the wave length of the pattern becomes $l_w=L/\sqrt{k^2+m^2}$.
Thus, for the parameters used in Fig.\ \ref{F5}, the random initial
conditions select the possible rotation angle
$\Psi=\pi/4$.
This choice corresponds to $m=k$ in Eq.\,(\ref{Psi}).
Next, the integer $m=4$ is chosen in such a way that the resulting wave length of the 
pattern,
$l_w=L/(m \sqrt{2})=3.53$, is 
close to the fastest growing wave length of $l_{\mathrm{max}}=3.43$.

For later use, we mention that
the spatial period $l_w$ and the angle $\Psi$
of a simulated periodic pattern
can be determined by computing the time-averaged power spectral density of the film thickness profile $h(x,y,t)$ according to
\begin{eqnarray}
\label{Spower}
  S_h({\bm k})= \frac{1}{T} \int_t^{t+T}|\hat{h}({\bm k})|^2\,dt,    
%
\end{eqnarray}
%
where 
$T$ is
the temporal period of oscillations
and
$\hat{h}({\bm k})$ denotes the discrete Fourier transform of $h(x,y,t)$.
%
%
The periodic boundary conditions for the square domain only allow for a discrete set of possible wave vectors forming a square
lattice with lattice constant
$\Delta k = 2\pi/L\approx 0.31$. 
Any periodic pattern in the height modulation $h(x,y)$ gives a
major peak of the power spectrum in Eq.\,(\ref{Spower}), which is
located at $k_x=2\pi m/L$, $k_y=2\pi k/L$, with the same integers $m$
and $k$ as in Eq.\,(\ref{Psi}).  Depending on the shape of the
surface, secondary peaks (higher harmonics) might be present,
but their strengths are typically orders of magnitude smaller compared
to the major peak.

In systems of active matter, stable square patterns have previously
been found in Vicsek-type models with memory. It was shown that in the
case of a ferromagnetic alignment between the self-propelled particles
with memory in the orientational ordering the system settles to a
perfectly symmetric state with a checkerboard arrangement of clockwise
and anti-clockwise vortices \cite{Aranson15}. Using our
classification, this checkerboard lattice corresponds to a square
pattern with the main axis tilted by $\Psi=\pi/4$ w.r.t. the
coordinate axes. Rectangular (nearly quadratic) positional
  order has also been reported as a state of collective dynamics in an
  active particle model with competing alignment interaction
  \cite{Grossmann14,GRBS2015epjt}.  A similar oscillation between
  stripe and square patterns has been reported for a mesoscopic
  continuum model for an active filament-molecular motor system where
  the oscillation is described as alternating wave between aster-like
  states that form a square lattice and stripe states \cite{Ziebert2006phd}. A
  related analysis of steady stripe and aster states is presented in
  \cite{ZiZi2005epje}.

%

Other experiments with active matter find hexagonal patterns.
For instance, a hexagonal lattice of vortices 
was observed in suspensions of highly concentrated spermatozoa of sea
urchins \cite{Riedel05}. Phenomenologically, the existence of
hexagonal patterns is often studied using a Swift-Hohenberg (SH)
equation for scalar fields \cite{CrHo1993rmp,Dunkel13_njp}. For such model
equations it is known that hexagonal structures can only be
stable if the model equations are not invariant under inversion of the
scalar field and that higher order gradient terms are needed
to stabilize square patterns \cite{Bestehorn2006}. In our case,
inversion symmetry is broken, i.e., Eqs.\,(\ref{smoluch}) and
(\ref{thin_film}) are not invariant under the simultaneous
transformations $h \rightarrow -h$ and $\rho \rightarrow -\rho$.
Nevertheless, in our numerical simulations we did not find stable
hexagonal patterns but find that square patterns dominate. 
This could imply  that higher order terms play an important role. Alternatively it
may indicate that a SH equation is not the appropriate order parameter
equation for our model that in contrast to standard variational SH equation 
has no gradient dynamics structure (see discussion in Section~\ref{Sec1} below Eq.~(\ref{thin_film})).

%
\subsection{Strongly perturbed square pattern} \label{sucsec.strongly}
%
Next, we study persisting patterns that emerge from the trivial state for parameters chosen far from the stability threshold. Thus, we set $V=3.5$, $D_{\rm eff}=600$, as in point $(4)$ in Fig.\,\ref{F2}(a). The steady state $h=1$, $\rho=1/(2\pi)$ is linearly unstable w.r.t.\ the finite wave number instability with a corresponding dispersion curve (not shown) similar to Fig.\,\ref{F2}(c). In 
the
square domain 
with
side length $L=20$, we 
start with the uniform state perturbed by
small-amplitude random 
noise.
After
a transient of about $20$ time units, 
a state evolves with an underlying square pattern, as demonstrated below, which is highly dynamic and
strongly perturbed by irregular temporal and spatial variations (see \href{https://www.dropbox.com/s/gipwhnd7xi6d4sm/V3d5_Deff_600.mp4?dl=0}{\color{red}{Movie 2}}).

\begin{figure} 
\centering
\includegraphics[width=0.95\textwidth]{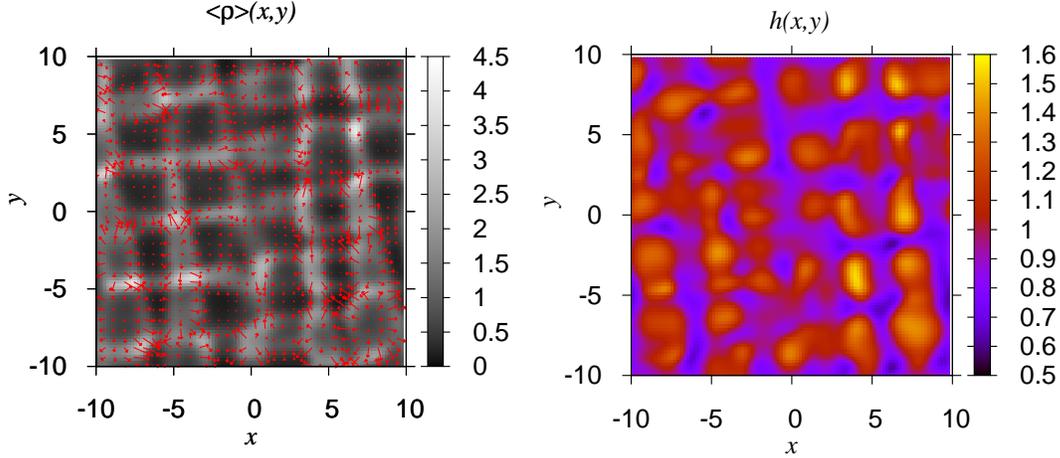}
\caption{(Color online) {Example of 
a strongly perturbed square pattern
at $V=3.5$ and $D=0.01$, i.e., $D_{\rm eff}=600$ for a domain size $L=20$. 
The temporal evolution started from the
homogeneous 
state with 
random
noise
added.
Snapshot 
is
taken at $t=100$.}
  \label{F8} }
\end{figure}

Snapshots of the pattern
in
Fig.\ \ref{F8} 
show the film thickness $h(x,y)$ and the average density $\langle\rho\rangle(x,y)$ at $t=100$. 
The latter varies between $\langle \rho \rangle_{\rm min}=0.1$ and $\langle \rho \rangle_{\rm max}=4.5$, in a much larger range
than for the 
regular pattern
in Fig.\ \ref{F5}.
One recognizes the underlying square pattern in the average density $\langle\rho\rangle(x,y)$ but the main lattice directions
are tilted against each other. 
The height profile $h(x,y)$ looks even stronger perturbed. Still
the elevated regions of the film surface (drops) are approximately arranged in a square lattice.
\href{https://www.dropbox.com/s/gipwhnd7xi6d4sm/V3d5_Deff_600.mp4?dl=0}{\color{red}{Movie 2}} shows how the tilted lattice planes in $\langle\rho\rangle(x,y)$ seem to split up and merge 
with their neighbors. The snapshot in Fig.\ \ref{F8} shows this scenario when going from left to right. This gives the
whole pattern a highly dynamic appearance.

\begin{figure}
\centering
\includegraphics[width=0.6\textwidth]{V3dot5_Deff_600_new.eps}
\includegraphics[width=0.39\textwidth]{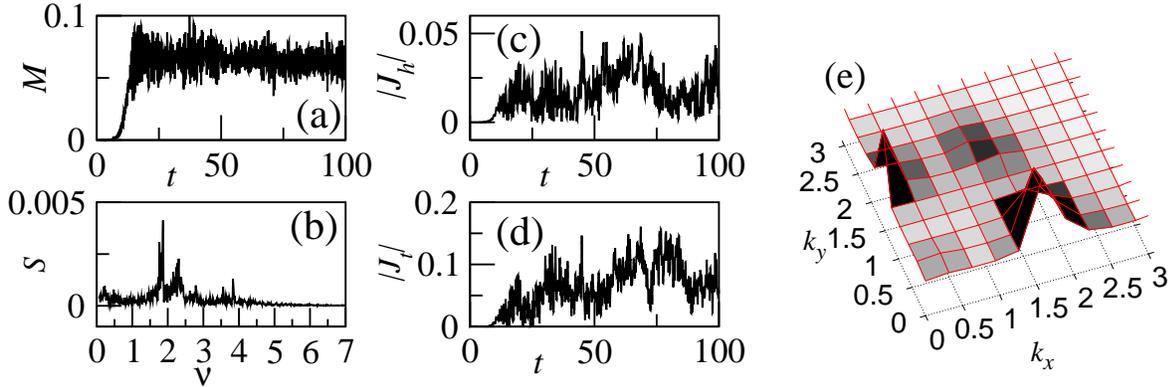}
\caption{(Color online) {
Temporal evolution
from the homogeneous 
state for parameters as in Fig.\ \ref{F8}. 
(a) The 
mode type $M$, 
(b) the spectral density $S(\nu)$
of $M(t)$ on the interval $t\in [20,100]$,
(c)
the magnitude of the fluid flux, $\mid \bar{J}_h\mid=\sqrt{(\bar J_h)_x^2+(\bar J_h)_y^2}$, 
(d) the magnitude of the translational flux of the swimmers, $\mid \bar{J}_t\mid=\sqrt{(\bar J_t)_x^2+(\bar J_t)_y^2}$, and
(e) 
the
time-averaged power spectral density 
of the height profile
obtained with Eq.\,(\ref{Spower}) by averaging over the interval $t\in [50,100]$.
}
  \label{F9}}
\end{figure}

The temporal evolution
of the pattern is visualized in Fig.\ \ref{F9}(a)-(d).  The mode type $M$
in plot (a)
is positive and oscillates randomly about its average value of $M\approx 0.08$. 
In Fig.\ \ref{F9}(b)
we plot the spectral density $S(\nu)$ of $M(t)$ calculated on the interval $t\in [20,100]$. 
We find a clear maximum at the frequency $\nu_{\mathrm{max}} \approx 1.87$
with small width $\Delta \nu / \nu_{\mathrm{max}} \approx 0.1$,
which corresponds to a period of $T = 1 / \nu_{\mathrm{max}} \approx 0.53$. 
The frequency $\nu_{\mathrm{max}}$ belongs to the pulsating pattern clearly recognizable in \href{https://www.dropbox.com/s/gipwhnd7xi6d4sm/V3d5_Deff_600.mp4?dl=0}{\color{red}{Movie 2}}.
A second, broader peak is
located at $\nu_{\mathrm{max}} \approx 2.3$ with 
width 
$\Delta \nu / \nu_{\mathrm{max}} \approx 0.2$. 
In addition, there exists a continuous background in $S(\nu)$, which
gives the pattern its random dynamic appearance.
%
Random oscillations of the magnitude of the fluid flux, $\mid \bar{J}_h\mid=\sqrt{(\bar J_h)_x^2+(\bar J_h)_y^2}$, 
[see Fig.\ \ref{F9}(c)]
and of the magnitude of the translational flux of the swimmers, $\mid \bar{J}_t\mid=\sqrt{(\bar J_t)_x^2+(\bar J_t)_y^2}$,
[see Fig.\ \ref{F9}(d)],
indicate global propagation of 
the pattern at each instance of time.  However, we find that the propagation direction
randomly changes with time 
with no preferred direction as expected for square symmetry.

%
%
%

%

Finally, to reveal the periodic structure of the pattern, we determined the
time-averaged power spectral density from Eq.\ (\ref{Spower}) averaged over the time interval $t\in [50,100]$.
As shown in Fig.\ \ref{F9}(e), the 
spectral
density has two major broad peaks: one is centered around $(k_x=2\pi m/L =1.57,\,k_y=2\pi k/L=0)$, i.e. $m=5$, $k=0$, 
and the other one is centered around  $(k_x=2\pi m/L=0,\, k_y=2\pi k/L =1.57)$, i.e. $m=0$ and $k=5$. 
These peaks correspond to a square pattern
with the main lattice directions
aligned along the coordinate axes, i.e. $\Psi= 0$ or $\Psi=\pi/2$. The dominating wave length 
or lattice constant
$l_w$ of the pattern
is 
$l_w = L/\sqrt{m^2+k^2}
=20/5
=4$. The third
peak with much less intensity at $(k_x=2\pi m/L=1.25,k_y=2\pi k/L=1.88)$, i.e. $m=4$ and $k=6$, 
can roughly be interpreted as a contribution from the sum of the two major wave vectors spanning the
reciprocal lattice.




\subsection{Multistability}
In order to systematically study the 
occurence and
stability of the 
two patterns studied in the previous sections,
we follow these patterns in parameter space using a primitive
``continuation method''. Namely, we take a snapshot of a converged
(time-dependent) state at some parameter value and use it as initial
condition for simulating the evolving pattern in a neighboring point
in parameter space.  The technique allows us to follow states, which
are linearly stable, and thereby identify multistability in parameter
space. Depending on the initial condition different stable spatio-temporal stable
patterns are obtained.
For an overview of proper continuation methods, which are also
  able to follow unstable steady states and therefore to determine the
  complete bifurcation diagram, see
Refs.~\cite{Kuznetsov2010,DWCD2014ccp}. However, these methods are not
readily available for time-periodic solutions of our PDE system.

\begin{figure}
\centering
\includegraphics[width=0.95\textwidth]{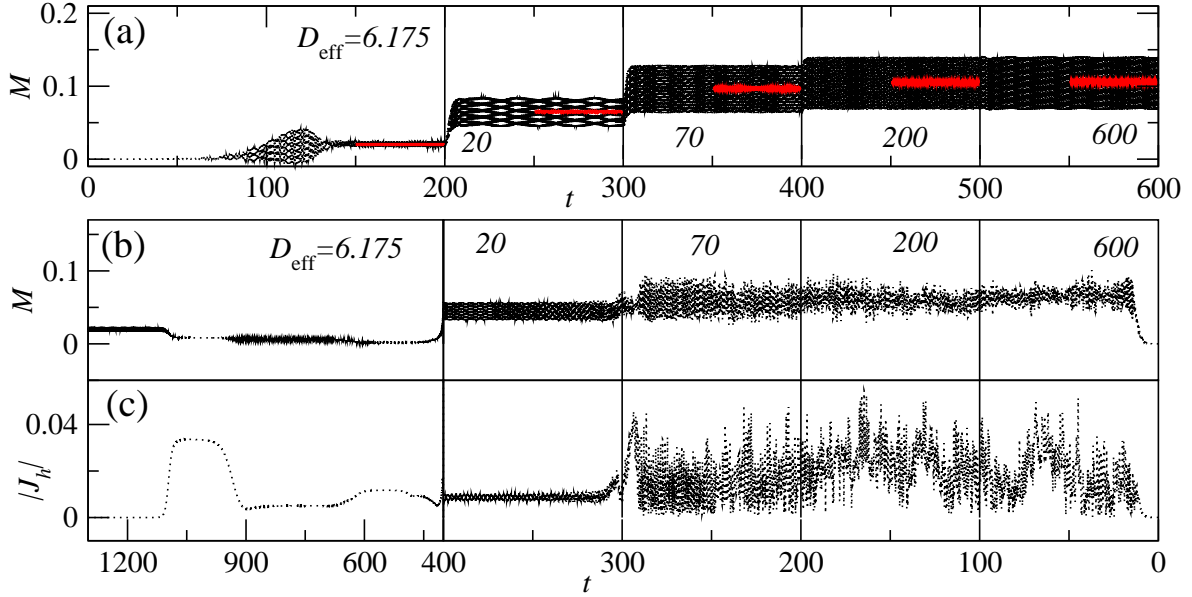}
\caption{(Color online) (a) The mode number $M$ during the continuation of the standing waves at fixed $V=3.5$ and $L=20$. Standing waves that emerged from the homogeneous steady state at $V=3.5$ and $D_{\rm eff}=6.175$ are numerically continued by increasing $D_{\rm eff}$. The vertical lines mark times when $D_{\rm eff}$ is changed. The respective values of $D_{\rm eff}$ for each interval are given by the numbers in each panel. The thick red solid line shows the mode type, averaged over one oscillation period. (b,c) Continuation of patterns with intermittent symmetry, emerged from the homogeneous steady state at $V=3.5$ and $D_{\rm eff}=600$ with $L=20$. Shown are (b) the mode number $M$ and (c) the modulus of the fluid flux $\sqrt{(\bar J_h)_x^2+(\bar J_h)_y^2}$ in dependence of time. The time axis is reversed.
  \label{F10}}
\end{figure}

First, we 
start with
the 
regular standing wave pattern, which we explored in Sec.\ \ref{subsec.standing} and
in Fig.\ \ref{F5} at fixed $V=3.5$.
We follow the standing wave solution
along the line connecting 
points $2$ and $4$ in Fig.\ \ref{F2}(a)
by gradually decreasing
$D$
(or increasing
$D_{\rm eff}$) in steps using four 
distinct values,
namely
$D_{\rm eff}=6.175\rightarrow 20 \rightarrow 70 \rightarrow 200 \rightarrow 600$. 
At each parameter point, we let the system settle into a stable state, which we identify by monotoring
mode type $M(t)$ and fluid flux $\bar{J}_h$.
The 
resulting
time evolution of 
mode type $M$ during the continuation schedule is shown in Fig.\,\ref{F10}(a). We find 
that the standing wave pattern keeps its main characteristics
up to the largest value
$D_{\rm eff}=600$ [point $4$ in Fig.\,\ref{F2}(a)]. 
In particular, mode type $M$ shows regular oscillations. The mean
value of $M$ [red solid line in Fig.\,\ref{F10}(a)]
and the oscillation amplitude of $M$ increase with $D_{\rm eff}$.
They reach their respective
maximal values of $0.1$ and $0.06$
at around $D_{\rm eff}=200$,
where the density and height variations are more pronounced compared to Fig.\ \ref{F5}.
Furthermore, the
oscillation period 
monotonically increases with $D_{\rm eff}$ (not shown). 
Interestingly, the
spatial period of the 
pattern
remains unchanged during the entire continuation schedule.

%
%

Next, we 
start with the strongly perturbed square pattern from Sec.\ \ref{sucsec.strongly} and Fig.\ref{F8} and follow the same path in Fig.\,\ref{F2}(a)
but this time backward from point 4 to point 2. For consistency, we use the same values of $D_{\rm eff}$ as 
in Fig.\,\ref{F10}(a).
The time evolution of the mode type $M$ and the modulus of the fluid flux $\mid \bar J_h\mid$ are 
plotted
in Figs.\ \ref{F10}(b) and (c) with 
reversed time axis. Remarkably, the system reaches the 
regular standing wave pattern only for parameters close to the
stability threshold, i.e., when $D_{\rm eff}$ is decreased to the
value in point $2$ in Fig.\ref{F2}(a).  After some transient dynamics,
visible in the time intervall from $t=400$ to 1100 in Figs.\
\ref{F10}(b) and (c), the system settles on the stable standing wave pattern.

\begin{figure}
\centering
\includegraphics[width=0.95\textwidth]{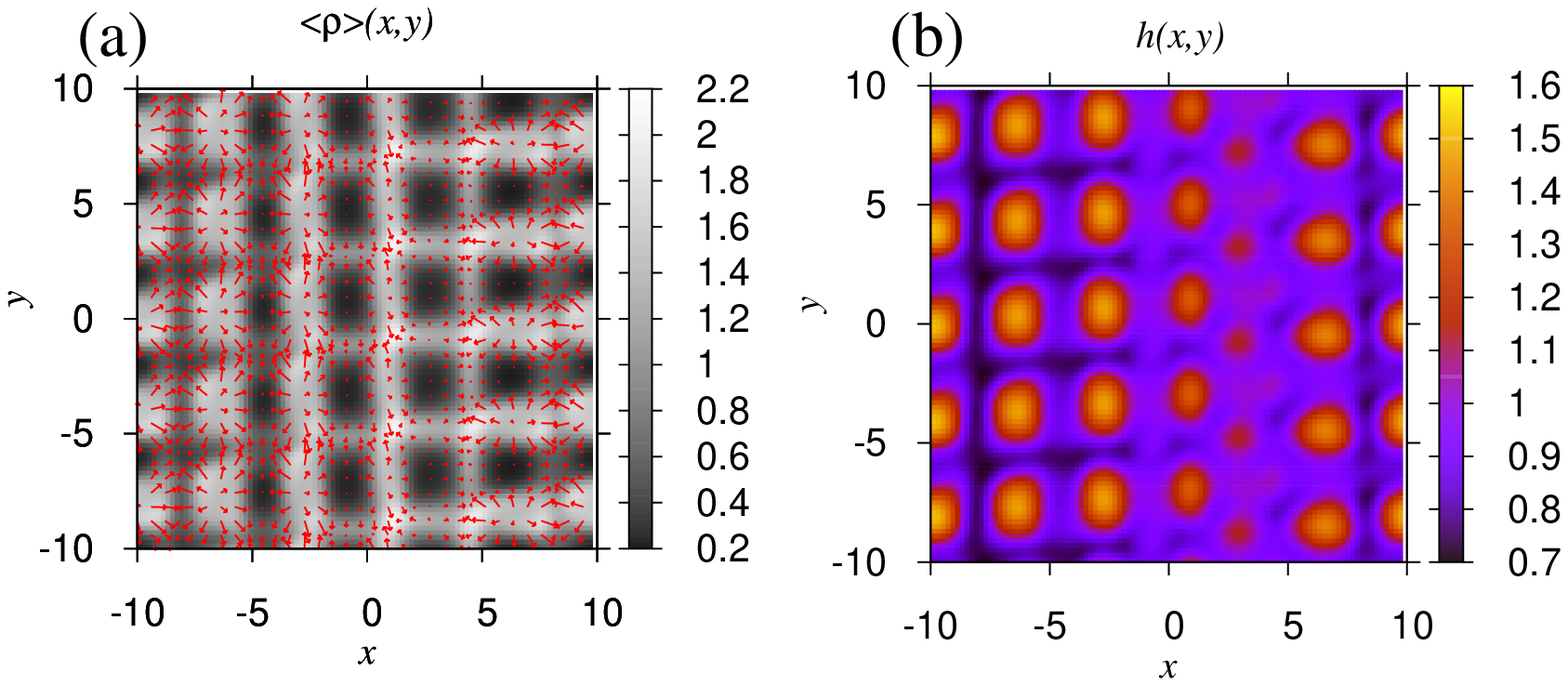}
\vspace{-1cm}

\includegraphics[width=0.55\textwidth]{V3dot5_Deff_20_mod.eps}
\includegraphics[width=0.3\textwidth]{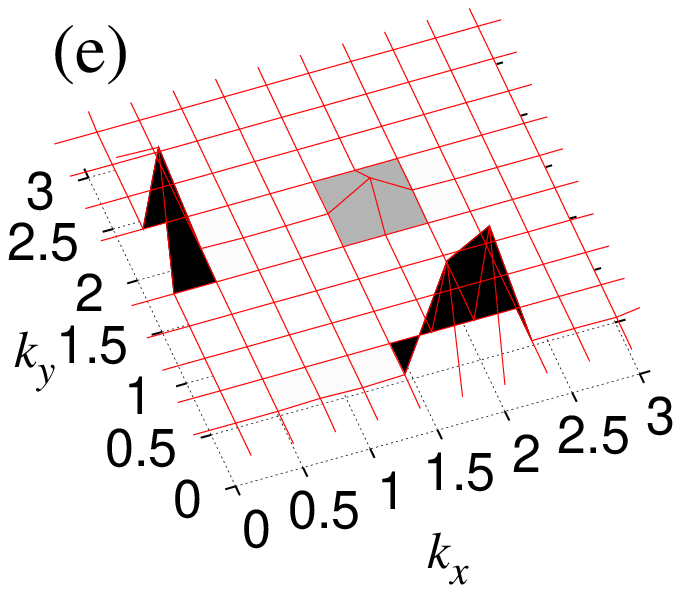}
\caption{(Color online) {(a,b) Persisting dynamic state at $V=3.5$ and $D_{\rm eff}=20$, during the numerical continuation of the patterns with intermittent symmetry in Fig.\ref{F10}. Snapshot taken at $t=400$. (c) The components of the translational surface flux $\bar{J}_t$. (d) the spectral density of $\mid \bar{J}_h\mid(t)$. (e)  Time-averaged power spectral density of the pattern, obtained with Eq.\,(\ref{Spower}) by averaging over the interval $t\in [350,400]$.
}
  \label{F11}}
\end{figure}

For larger values of $D_{\rm eff}$, i.e., further away from the threshold, we found that the system is multistable.
At $D_{\rm eff} > 20$ stable regular standing waves still exist but also patterns similar to the strongly perturbed square pattern as
shown in Fig.\ref{F8} are stable. However, at $D_{\rm eff}=20$, the 
dynamics of the square pattern becomes more regular but still keeps the feature of lattice planes splitting and merging
with their neighbors. This is demonstrated by \href{https://www.dropbox.com/s/rlat7yr5fwtx8au/V3d5_Deff_20.mp4?dl=0}{\color{red}{Movie 3}} and by the snapshots
in Fig.\ \ref{F11}(a) and (b), taken at $t=400$ during the numerical continuation in Fig.\ \ref{F10}. 

%
%
%
Figure\ \ref{F11}(d) shows the $x$ and $y$ components of the translational flux of the swimmers. One clearly recognizes 
directed motion,
on average, into the negative $x$ and positive $y$ direction, which is also visible in \href{https://www.dropbox.com/s/rlat7yr5fwtx8au/V3d5_Deff_20.mp4?dl=0}{\color{red}{Movie 3}}. The components of the fluid flux 
$\bar{J}_h(t)$ behave similarly. Furthermore, both flux components show a fast oscillation with a weak slow modulation superimposed.
By taking the Fourier transforms of $\mid \bar{J}_h (t) \mid$ [Fig.\ \ref{F11}(d)], one identifies
a dominant peak at $\nu = 3.5 $ corresponding to a period of $T=0.28$ of the fast oscillations. They result from the pulsation in 
the square pattern as \href{https://www.dropbox.com/s/rlat7yr5fwtx8au/V3d5_Deff_20.mp4?dl=0}{\color{red}{Movie 3}} demonstrates, in  particular, for the height profile. The weak modulation generates a small peak in 
the power spectrum with frequency $\nu = 0.36$ or period $T=2.8$. It is not really recognizable in the time evolution of \href{https://www.dropbox.com/s/rlat7yr5fwtx8au/V3d5_Deff_20.mp4?dl=0}{\color{red}{Movie 3}}.
Otherwise, the continuous part of the spectrum as observed in Fig.\ \ref{F9}(b) for the strongly perturbed square pattern is
missing here since the square pattern has a more regular dynamics.

Finally, the
time-averaged power spectral density $S_h$, averaged over the interval $t\in [350,400]$, is given in Fig.\ \ref{F11}(e). 
The two
major peaks correspond to the square pattern with spatial period of $L/6$ aligned along the coordinate axes. 
The third,
much weaker
peak at $k_x=k_y=2\pi m/L = 1.57$ with $m=5$ again roughly corresponds to a contribution of the two major wave vectors spanning 
the reciprocal lattice.


%
%
%

Multistability of several persisting dynamic states under identical external conditions was also found in other systems. 
For example, in experiments on groups of schooling fish \cite{Couzin13} it
was observed that depending on the starting conditions and/or the nature of perturbations, as well as the group size, the fish group may exhibit two different dynamic states: the so-called milling state, which is characterized by fish swimming in a large circle, and the polarised state, which corresponds to fish swimming predominantly in one direction.

\subsection{Persisting traveling patterns}
%
\begin{figure}
\centering
\includegraphics[width=0.95\textwidth]{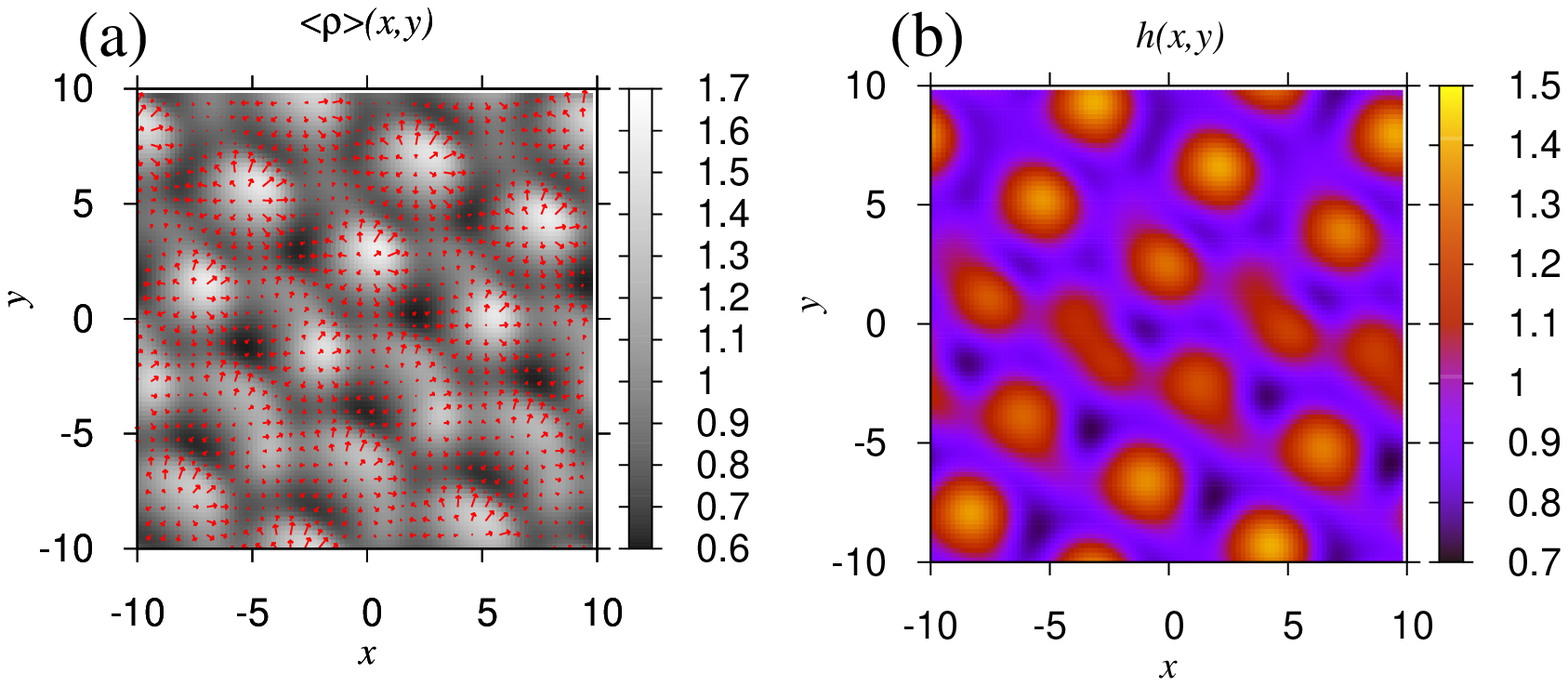}
\vspace{-0.1cm}
\includegraphics[width=0.5\textwidth]{V2_D2d4_new.eps}
\includegraphics[width=0.4\textwidth]{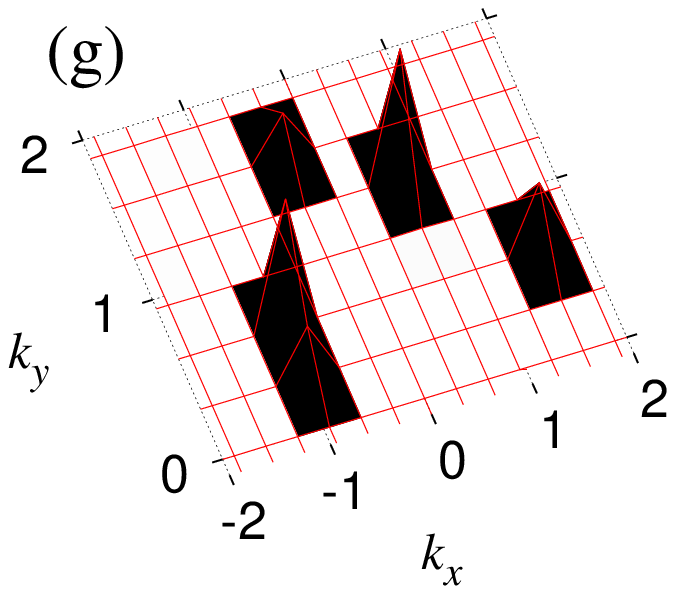}
\caption{(Color online) {(a,b) Snapshots of the persisting traveling
    pattern, which emerges from the mixed instability type for $V=2$,
    $D=2.4$, ($D_{\rm eff}=0.88$), $L=20$ [see point $3$ in Fig.\
    \ref{F2}(a)]. The corresponding dispersion curve is shown in Fig.\
    \ref{F2}(d).  (c) Mode type $M$, (d) $x$ and $y$ component of the
    space averaged fluid flux $\bar {\bm J}_h$,
(e) the film height $h(t)$ at a randomly chosen point on the film surface, and (f)
its power spectrum $S_h$.
(g) Power spectrum for the spatial modulation of the height profile [see Eq.\ (\ref{Spower})] averaged over the interval $t\in [800,900]$.
}
  \label{F12}}
\end{figure}

In the triangular parameter region in Fig.~\ref{F2}, where the mixed instability occurs, one finds persisting patterns, i.e., long-time stable spatio-temporal patterns that are characterized by a time-independent mode type and a constant non-zero fluid flux, i.e., travelling waves.
Figures\ \ref{F12}(a) and (b)  give 
example snapshots for such a pattern obtained at point $3$ in Fig.\,\ref{F2}(a) at system size $L=20$ and at 
time $t=900$. \href{https://www.dropbox.com/s/tba6mi7rjvfkjlo/V2_Deff_0d88.mp4?dl=0}{\color{red}{Movie 4}}  
reveals a pattern traveling approximately along the diagonal with a propagation speed estimated to be of the order of the self-propulsion 
velocity $V=2$. The maxima in the height profile form a rectangular lattice. Perturbations run
along the lattice lines and shift the maxima by roughly half a lattice constant; presumably, to match the periodic boundary condition
for the whole square domain.
The time evolutions of mode type $M$
and of the components of the fluid flux, $(J_{h})_x$ and $(J_{h})_y$,
are shown in Fig.\ \ref{F12}(c)
and (d),
respectively. 
The height modulation at an arbitrary point plotted in Fig.\ \ref{F12}(e) looks rather irregular. Its power spectrum in Fig.\ \ref{F12}(f)
reveals one peak at $\nu = 0.32$, which corresponds to one shift motion of a bump in the height profile. Frequencies at $\nu = 0.1$ and $0.17$
belong to longer cycles of two or three shifts. The power spectral density for the spatial modulations in the film height,
averaged over the interval $t\in [800,900]$, 
is plotted in Fig.\ \ref{F12}(g). In the upper and the lower halves of the snapshot in Fig.\ \ref{F12}(b) one can clearly see a rectangular pattern with the aspect ratio of $\approx 1.3$. Remarkably, the aspect ratio of $\approx 1.3$ of the rectangular pattern is not compatible with the periodic boundary conditions of the square domain. Nontheless, the pattern is fitted into the square domain due to the presence of a defect-like modulation, seen at the center of the snapshot. The defect disturbes the rectangular lattice dynamically, as it continuously moves along the lattice lines and shifts the elevations in the height profile (see  \href{https://www.dropbox.com/s/tba6mi7rjvfkjlo/V2_Deff_0d88.mp4?dl=0}{\color{red}{Movie 4}}). 
As a
result, in the power spectrum in Fig.\ \ref{F12}(g) we find two major peaks located at $\{k_x^{(1)}=0.62,k_y^{(1)}=1.25\}$ and $\{k_x^{(2)}=-0.94,k_y^{(2)}=0.62\}$ and one smaller peak located at $\{k_x^{(3)}=-0.94,k_y^{(3)}=0.31\}$. 
The remaining two weaker peaks are again higher contributions from wave vectors in the reciprocal lattice.

Persisting states characterized by propagating structures are well known for active matter systems. Thus, traveling density waves were found in experiments with an assay of actin filaments, driven by motor proteins \cite{Schaller10}. At the leading edge (lamellipodium) of a crawling cell, the alignment of actin filaments along the substrate leads to the forward translation of lamellipodium and thus, to cell motility \cite{Small02}. Moving density stripes and propagating isolated density clusters have been found in microscopic Vicsek-type models and in continuum models of self-propelled particles \cite{Chate08,Gopinath12,Mishra10}.

\subsection{Random patterns}
\label{patt-random}

\begin{figure}
\centering
\includegraphics[width=0.95\textwidth]{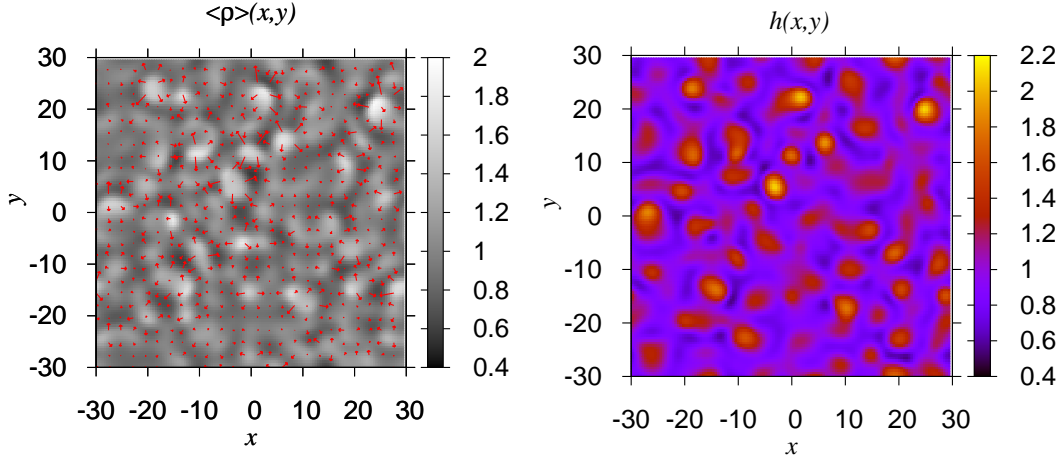}
\caption{(Color online) {
Random
patterns 
emerge
in the region of the
zero wave number 
instability
at $V=3.5$ and $D_\mathrm{eff}=0.66$ in 
a
system of 
size $L=60$. Number of the Fourier modes for the discretisation in space is $N=128$.}
  \label{F13}}
\end{figure}

When studying the nonlinear behaviour for $V=3.5$ and $D_{\rm
eff}=0.66$ [point $1$ in Fig.\,\ref{F2}(a)], we find truly random
patterns emerging from the zero wave-number instability. The
corresponding dispersion curve is shown in Fig.\,\ref{F2}(b) and gives
the fastest growing wave length as $l_\mathrm{max}\approx 9 $. We set
$L=60$, $N=128$ and start the simulations at the trivial state. After
a transient, the system settles to an irregular pattern in space and
time that oscillates randomly and locally travels in random directions.  
Typical snapshots are shown in Fig.\ \ref{F13} and  \href{https://www.dropbox.com/s/qb4mkxrx7gpzdjc/V3d5_Deff_0d66.mp4?dl=0}{\color{red}{Movie 5}} illustrates the irregular spatio-temporal
pattern.

\begin{figure}
\centering
\includegraphics[width=0.3\textwidth]{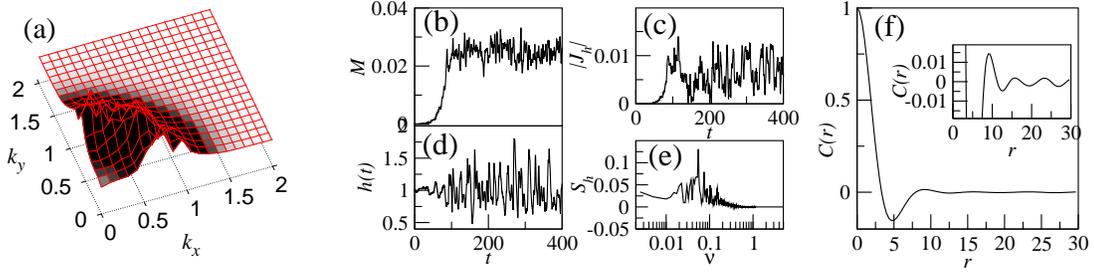}
\includegraphics[width=0.6\textwidth]{correlation_new.eps}
\caption{(Color online) {
(a) Time-averaged spectral density from Eq.\,(\ref{Spower}) computed numerically by averaging the Fourier transformed film thickness patterns in Fig.\ref{F13} over $t\in[150,350]$. (b,c) The mode type $M$ and the modulus of the fluid flux $\mid \bar{J}_h\mid$ over time. (d) Time evolution of $h(t)$ at a randomly chosen point on the film surface. (e) Spectral density $S_h$ of $h(t)$ from (d). (f) Normalised radial autocorrelation function $C(r)$. Inset shows zoomed graph of $C(r)$ around zero.
}
  \label{F14}}
\end{figure}

The randomness of the pattern is clearly visible in the time evolution of mode type $M$ and of the modulus of the fluid flux $\mid \bar{J}_h\mid$,
which we plot in Figs.\ \ref{F14}(a) and (b), respectively. In Fig.\ \ref{F14}(a) the power spectral density of the height profile averaged over the 
interval $t\in [150,350]$ reveals a clear maximum at $\mid {\bm k}\mid = 0.7$. The spectral density is radially symmetric, which implies that spatial 
correlations in the pattern only depend on the distance $r$ between any two points on the film surface. Otherwise, it is continuous 
as expected for a random pattern. Via the convolution theorem, the time-averaged power spectral density from Eq.\ (\ref{Spower}) 
is directly related to the spatial height correlation function 
\begin{eqnarray}
\label{correl}
C({\bm r}) = \frac{C_0}{t_2-t_1}\int_{t_1}^{t_2} dt\,\int [h({\bm r}+{\bm w})-1][h({\bm w})-1]\,d{\bm w},
\end{eqnarray}
%
where $t_1=150$ and $t_2=350$ and the constant $C_0$ is chosen such that $C(0)=1$. We plot the normalised radially-symmetric 
correlation function $C(r)$ in Fig.\ \ref{F14}(f) over the length of half the system size, i.e., $L/2=30$. $C(r)$ rapidly decreases 
with $r$ and drops by two orders of magnitude over the distance of $r=10$, as shown in the inset of Fig.\ \ref{F14}(f). Correlations become
negligibly small at distances larger than 10 and the pattern looks random. The oscillation at small distances corresponds to the maximum in 
the power spectral density. They are caused by wave fronts traveling in random directions, which one recognizes in the orientation-averaged 
swimmer density in \href{https://www.dropbox.com/s/qb4mkxrx7gpzdjc/V3d5_Deff_0d66.mp4?dl=0}{\color{red}{Movie 5}}. 

Finally, the film height at an arbitrary position changes randomly in time, as shown in  Fig.~\ref{F14}(d). 
However, the spectral density $S_h$ plotted in Fig.~\ref{F14}(e) has a clear peak at $\nu =0.055$. This implies that the temporal dynamics of the patterns cannot be regarded as purely random.

Highly dynamic random spatio-temporal patterns of active matter are known as quasi or mesoscale turbulence. Irregular turbulent states have been found in experiments with dense bacterial suspensions \cite{Dunkel13_prl,Goldstein14,sokolov07,sokolov09,liu12,sokolov09_pre,wensink_pnas,ishikawa11} and in active microtubuli networks\cite{Sanchez12}.

\section{Discussion and Conclusion}

We have investigated the collective behaviour of a colony of point-like non-interacting self-propelled particles (microswimmers) 
that swim at the free surface of a thin liquid layer on a solid support. In contrast to former work \cite{AlMi2009pre}, where the 
motion of the particles was considered to be purely orthogonal to the free surface, here we have 
also
allowed for 
active
motion parallel to the film surface. The resulting coupled dynamics of the swimmer density $\rho(x,y,\phi,t)$ and the film thickness 
profile $h(x,y,t)$ is captured in a long-wave model in the form of a Smoluchowski equation for the one-particle density 
$\rho(x,y,\phi,t)$ and a thin film equation for $h(x,y,t)$ that allows for (i) diffusive and convective transport of the swimmers 
(including rotational diffusion), (ii) capillarity effects (Laplace pressure) including a Marangoni force caused by gradients in the 
swimmer density, (iii) and a vertical pushing force of the swimmers that acts onto the liquid-gas interface. 

First, we have extended the linear stability analysis of the homogeneous and isotropic state that was presented before in Ref.~\cite{PoTS2014pre} focusing, in particular, on the characteristics of the two instability modes (one 
at
zero wave number and one 
at
finite wave number) and their 
mixed appearance
close to the border of the stable region in the 
stability diagram
spanned by the swimmer speed $V$ and the effective diffusion constant $D_\mathrm{eff}$. 

Our linear stability 
analysis indicates
that the onset 
and dispersion relation of the zero-wave number instability mode do not fit well into the classification scheme of Cross and Hohenberg \cite{CrHo1993rmp}. 
The
long-wave instability of the free film surface
occurs at
$k=0$,
where
the zero-wave number mode has zero imaginary part. However, arbitrarily close to onset in the unstable region, the fastest growing wave number $k_{\rm max}$ corresponds to a pair of complex conjugate eigenvalues. Moreover, 
the entire unstable band of wave numbers, $0<k<k_c$, does \textit{always} contain a range of small $k<k_\mathrm{rc}\sim|\eta|$
(with $\eta$ measuring the distance from the stability threshold), where the first two leading eigenvalues are real (one $\sim k^2$ and one $\sim k^4$), and a range $k_{\rm rc}<k<k_c$, where the two leading eigenvalues form a complex conjugate pair.  The latter range always contains the fastest growing wave number $k_\mathrm{max}$. 
This indicates that 
this
zero-wave number instability is similar to a zero-frequency Hopf bifurcation in dynamical systems \cite{Kuznetsov2010} and in the context of the instabilities of spatially extended systems, it might be called a zero-frequency type $II_o$ instability.

The behaviour at about $k=0$ has also important implications for a weakly nonlinear theory for the short-wave instability at $k=k_c$.
Such a theory would need to take into account that the slow complex modes around $k=k_c$ 
couple to 
the
two 
unstable long-wave modes at $k\approx0$
with real eigenvalues.
We believe that such a coupling is responsible for the observed wave behaviour, where a travelling wave is perturbed by a long-wave modulation, as in Fig.\ref{F11}. 
The two long-wave modes are a direct consequence of 
the existing
two conserved quantities in the system: the mean film height and the orientation-averaged mean 
swimmer
concentration. 
Simpler cases with one long-wave mode (resulting from a single conserved quantity) that couples to a short-wave mode have been considered in Refs.~\cite{MaCo2000n,CoMa2003pd,WiMC2005n}.
Such an analysis is not feasible in our case, where the evolution equations capture the dynamics in two spatial dimensions \textit{and} account for a fully $\phi$-dependent density $\rho(x,y,\phi,t)$.
However, a 
one-dimensional 
model
system, where instead of rotational diffusion the swimmers can only flip between 
swimming to the left or right
shows similar transitions and lends itself to a
weakly nonlinear analysis. Such a simplified system is under investigation and will be presented elsewhere.

Numerical simulations of the time evolution equations 
(\ref{smoluch}) and (\ref{thin_film})
reveal a rich variety of persisting dynamic states. 
We
have abstained from a comprehensive 
parameter study
of the different persisting states. Instead, we have given an overview of the zoo of dynamic states that can be found,
when solving
the system of equations\ (\ref{smoluch}) and (\ref{thin_film}).

In particular, for parameters chosen in the vicinity of the stability threshold of the finite wave number instability, we have found 
a
highly regular 
dynamic
standing 
wave pattern
by starting the simulations with small 
random perturbations of the trivial state. The 
standing wave
pattern
is characterised by a regular array of elevations of the free 
film
surface 
that periodically transform and rearrange following a rather complex pathway. Thus, over one quarter of the oscillation cycle, the elevations transform from a perfect square lattice into an array of stripes followed by 
a new square lattice of elevations that is shifted w.r.t.\ the initial square lattice by 
exactly one half of the spatial 
wave
period.
The transformation of the average density of swimmers follows the pattern of the film thickness profile. The swimmers are arranged
in a regular lattice with high and low density spots. Spatial variations of the film thickness profile and the 
orientation-averaged
density profile are in-phase implying that high density spots sit approximately on top of the droplets. The orientation of swimmers 
in each high density spot shows strong polar order
with a hedgehog defect right at the maximum. 
The space-averaged fluid flux of the square wave is zero at all times. 

Next, we have employed a 'primitive' continuation method and followed the 
standing 
wave patterns
through parameter space moving further away from the stability threshold. Our numerical results suggest that 
standing 
wave patterns
exist and are stable possibly in the entire region, where the homogeneous state is linearly unstable w.r.t.\ the finite wave number 
instability (i.e. for $D_{\rm eff}>1$). 
On the reverse path, initial random perturbations of the homogeneous and isotropic state develop into a strongly perturbed
square pattern, where lattice lines continuously split and merge with their neighbors.
Unlike for the 
regular
standing 
wave pattern,
the average fluid flux 
oscillates randomly about zero. 
However, the time-averaged
power 
spectral
density of the 
height
profile reveals 
a clear square pattern.
Further decreasing $D_{\rm eff}$ towards the stability threshold, the dynamics of the perturbed square lattice 
becomes regular. The fluid flux in this state is periodic in time with a weak modulation
superimposed. The major frequency corresponds to a pulsation of the square pattern.


By choosing the parameters in the mixed-instability region,
we find 
a persisting traveling pattern
characterised by 
constant
space-averaged fluid flux. 
Elevations in the film surface are
arranged in a 
rectangular
lattice that travels in 
one
direction with the speed of the order of the self-propulsion velocity,
while perturbations continuously shift the hight elevations. 
At their positions the
swimmers form
high-density spots with strong polar order
around a hedgehog defect
similar to the 
regular
standing wave. Finally, choosing parameters from the region
with
the zero wave number instability, one finds a 
random spatio-temporal pattern
with a 
correlation length much smaller than the system size.

Our findings clearly show that a rich variety of persisting regular and irregular dynamic states can be found in 
an
active matter 
system
of 
self-propelled particles
without direct interactions.
In our model, the interaction between the swimmers occurs on 
a
coarse-grained level
and is
mediated by 
large-scale deformations of the liquid film.
Similar
types of 
dynamic states found here 
were
previously observed in other active matter systems of interacting particles as indicated at the respective ends of 
sections \ref{subsec.standing} to \ref{patt-random}. For instance, stable square patterns 
were
found in Vicsek-type models with memory \cite{Aranson15}. Multistability of the system under identical external conditions 
was reported earlier in experiments with
groups of schooling fish \cite{Couzin13}. Various traveling states 
occurred
in experiments with 
motility
assays of actin filaments driven by motor proteins \cite{Schaller10}, in microscopic Vicsek-type models, and in continuum models of 
self-propelled particles \cite{Chate08,Gopinath12,Mishra10}. 
Finally, random or
turbulent states 
were
observed in experiments with dense bacterial suspensions \cite{Dunkel13_prl,Goldstein14,sokolov07,sokolov09,liu12,sokolov09_pre,wensink_pnas,ishikawa11} and in active microtubule 
networks 
\cite{Sanchez12}. 
It is fascinating that active particles acting as surfactants at the surface of a thin liquid film provide a model system, where
all these different dynamic patterns can be realized by tuning appropriate parameters.




Possible extensions of the model include the incorporation of wettability effects 
by adding the
Derjaguin (disjoining) pressure to study swimmer carpets not only on films but also on shallow droplets 
and,
in particular, 
interactions with (moving) contact lines. One may also go beyond the approximation of point-like non-interacting particles by introducing finite size effects (short-range interactions between particles) as well as long-range
(hydrodynamic)
interactions. The resulting Smoluchowski equation would 
then contain non-local terms as in dynamical density functional theories for the diffusive dynamics of interacting colloids, 
polymers, and macromolecules \cite{MaTa1999jcp,ArEv2004jcp}. For a consistent model also the film height equation would 
require
additional terms that may be determined via the gradient dynamics formulation \cite{ThAP2012pf} that has to be recovered 
in the limit of passive surfactant particles/molecules.

\section{Appendix: Semi-implicit numerical scheme for Eqs.\,(\ref{thin_film},\ref{smoluch})}
\label{AppendixA}
In the employed dimensionless quantities, the resulting coupled system consists of the reduced Smoluchowski equation and the thin film equation. It reads
\begin{eqnarray}
\label{nondim}
\partial_t h&+&{\bm \nabla}\cdot {\bm J}_h=0,\nonumber\\
\partial_t \rho &+& {\bm \nabla}\cdot {\bm J}_{\rm t}+ \partial_\phi J_\phi=0,
\end{eqnarray}
with the dimensionless fluid flux ${\bm J}_h$, the translational and rotational probability currents ${\bm J}_t$ and $J_\phi$, the surface fluid velocity ${\bm U}_\parallel$ and the vorticity of the fluid flow $\Omega_z$ 
\begin{eqnarray}
\label{fluxes}
{\bm J}_h&=&\frac{h^3}{3}{\bm \nabla}\left[\Delta h + \beta \langle \rho \rangle \right] -\frac{1}{2}\left(h^2 {\bm \nabla \langle \rho\rangle}\right),\nonumber\\
{\bm J}_{\rm t}&=&\left(V{\bm q}+{\bm U}_\parallel -d{\bm \nabla}\right)\rho,\nonumber \\
J_\phi&=&\frac{1}{2}\Omega_z\rho- D \partial_\phi \rho,\nonumber\\
{\bm U}_\parallel&=&-h{\bm \nabla}\langle \rho\rangle + \frac{h^2}{2}{\bm \nabla}\left( \Delta h + \beta\langle \rho \rangle \right),\nonumber\\
\Omega_z&=&\partial_x U_y -\partial_y U_x.
\label{eq.several}
\end{eqnarray}
The coupled Eqs.\,(\ref{nondim}) are solved numerically using the following version of the semi-implicit spectral method. First, we average the density equation over the orientation angle $\phi$. This yields
\begin{eqnarray}
\label{ap_eq1}
\partial_t \langle \rho\rangle  &+& {\bm \nabla}\cdot \langle {\bm J}_{\rm trans}\rangle=0,
\end{eqnarray}
with the average translational current $\langle {\bm J}_{\rm trans}\rangle = V \langle{\bm q}\rho\rangle + ({\bm U} -d{\bm \nabla})\langle \rho \rangle$ and ${\bm q}=(\cos{\phi},\sin{\phi})$. 
It is worthwhile noticing that the only term in Eq.\,(\ref{ap_eq1}) that depends on the three-dimensional density $\rho(x,y,\phi,t)$ is the average orientation vector $\langle{\bm q}\rho\rangle$. All other terms in Eq.\,(\ref{ap_eq1}), including the surface fluid velocity ${\bm U}$ explicitly depend on the average density $\langle \rho \rangle$.

Next, we group the thin film equation together with Eq.\,(\ref{ap_eq1}) 
\begin{eqnarray}
\label{ap_eq2}
\partial_t h &+&{\bm \nabla}\cdot {\bm J}_h=0,\nonumber\\
\partial_t \langle \rho\rangle  &+& {\bm \nabla}\cdot \langle {\bm J}_{\rm trans}\rangle=0,
\end{eqnarray}
with the fluid flux ${\bm J}_h = \frac{h^3}{3}{\bm \nabla}\left[\Delta h + \beta \langle \rho \rangle \right] -\frac{1}{2} {\bm \nabla}\left(h^2 {\bm \nabla \langle \rho\rangle}\right)$.

At the next step, we single out the linear parts in all the terms in Eqs.\,(\ref{ap_eq2}) that explicitly depend on the average density $\langle \rho \rangle$. This is done by linearising the current ${\bm J}_t$ and the fluid flux ${\bm J}_h$ about the trivial steady state given by $h=1$ and $\langle \rho\rangle =1$. 

Finally, following the standard implicit time-integration scheme, we replace $\partial_t h $ and $\partial_t \langle \rho \rangle $  by $(h^{t+dt}-h^t)/dt$ and by  $(\langle \rho \rangle^{t+dt}-\langle \rho \rangle^t)/dt$, respectively and take all linear terms at time $t+dt$ and all nonlinear terms, including the term $V\langle{\bm q}\rho\rangle$, at time $t$. Upon these transformations Eqs.\,(\ref{ap_eq2}) become
\begin{eqnarray}
\label{ap_eq3}
\frac{h^{t+dt}-h^t}{dt} &+&\frac{1}{3}\Delta^2 h^{t+dt}+\left(\frac{\beta}{3}-\frac{1}{2}\right)\Delta \langle \rho \rangle^{t+dt}+{\bm \nabla}\cdot ({\bm NL}_h)^{t}=0,\nonumber\\
\frac{\langle \rho \rangle^{t+dt}-\langle \rho \rangle^t}{dt} &+& \frac{1}{2}\Delta^2 h^{t+dt}+\left(\frac{\beta}{2}-1-d\right)\Delta \langle \rho \rangle^{t+dt}+{\bm \nabla}\cdot (\langle {\bm NL}_{\rm trans}\rangle)^t=0,
\end{eqnarray}
with the nonlinear parts given by 
\begin{eqnarray}
\label{ap_eq4}
({\bm NL}_h)^{t}& =& \left[\frac{h^3-1}{3}{\bm \nabla}\left[\Delta h + \beta \langle \rho \rangle \right] -\frac{1}{2} {\bm \nabla}\left([h^2-1] {\bm \nabla \langle \rho\rangle}\right)\right]^t,\nonumber \\
(\langle {\bm NL}_{\rm trans}\rangle)^t &=&\left[V \langle{\bm q}\rho\rangle + ({\bm U})\langle \rho \rangle -\frac{1}{2}{\bm \nabla}(\Delta h)-\left(\frac{\beta}{2}-1\right){\bm \nabla}\langle \rho \rangle\right]^t.
\end{eqnarray}
After taking the discrete Fourier transforms of Eqs.\,(\ref{ap_eq3}), we find the updated fields $h^{t+dt}$ and $\langle \rho \rangle^{t+dt}$ at the time step $t+dt$.

With the update average density $\langle \rho \rangle^{t+dt}$ and the film thickness $h^{t+dt}$ at hand, we find the updated surface fluid velocity ${\bm U}^{t+dt}$ and the updated vorticity $\Omega^{t+dt}$. These functions are then substituted into the three-dimensional density equation
\begin{eqnarray}
\label{ap_eq5}
\frac{\rho^{t+dt}-\rho^t }{dt} &+& {\bm \nabla}\cdot {\bm J}_{\rm trans} + \partial_\phi J_{\rm rot}=0,
\end{eqnarray}
with the translational current ${\bm J}_{\rm trans} = V ({\bm q}\rho)^t + ({\bm U}^{t+dt})\rho^t -d{\bm \nabla}\rho^{t+dt}$ and the rotational current $J_\phi = (1/2)\Omega^{t+dt}\rho^t - D(\partial_\phi \rho^{t+dt})$.

After taking the Fourier transform of Eq.\,(\ref{ap_eq5}) both, in space as well as in the angle $\phi$, we find the updated three-dimensional density $\rho^{t+dt}$.
%


\begin{thebibliography}{74}
\bibitem[{not()}]{note_omega}
\bibinfo{note}{Note that $\Omega_z$ can also be introduced as
  $\lim_{z\rightarrow h} ({\bm \nabla} \times {\bm u})_z$, where $({\bm \nabla}
  \times {\bm u})_z$ denotes the vertical component of the fluid vorticity in
  the bulk of the film. The difference between $\lim_{z\rightarrow h} ({\bm
  \nabla} \times {\bm u})_z$ and $\partial_x u_y-\partial_y u_x$ with
  $(u_x,u_y)$ from Eq.(\ref{surface_vel}) is of the second order in parameter
  $\epsilon$ of the long-wave approximation, and thus, does not effect the
  results of the linear stability of the homogeneous steady state. However,
  certain aspects of the nonlinear evolution may vary depending on the
  definition of $\Omega_z$}.
\end{thebibliography}

\end{document}